\documentclass[a4paper,12pt]{article}  % option 'draft' for faster compiling; switch off to see images
\pdfoutput=1
\usepackage{amsmath,amsfonts,amssymb,epsfig,comment,xspace,listings,hyperref,cite}
\usepackage{graphicx,caption,subcaption}

\numberwithin{equation}{section}
\usepackage{slashed}

\usepackage{youngtab}
\usepackage{listings}
\usepackage{xcolor}
\usepackage{booktabs}

\usepackage{tabularx, multirow}

\makeatletter

\def\pd[#1]{\frac{\partial}{\partial #1}}
\def\pdd[#1,#2]{\frac{\partial #1}{\partial #2}}

\def\beq{\begin{equation}}
\def\eeq{\end{equation}}

\def\twomat[#1,#2][#3,#4]{\left( \begin{array}{cc} #1 & #2 \\ #3 & #4 \end{array} \right)}
\def\twoa[#1,#2][#3,#4]{\left( \begin{array}{cc} #1 & #2 \\ #3 & #4 \end{array} \right)}

\def\thv[#1,#2,#3]{\left( \begin{array}{c} #1 \\ #2 \\ #3 \end{array} \right)}
\def\twv[#1,#2]{\left( \begin{array}{c} #1 \\ #2 \end{array} \right)}

\def\SARAH{{\sc SARAH}\xspace}
\def\SPheno{{\sc SPheno}\xspace}
\def\MicrOMEGAs{{\sc MicrOMEGAs}\xspace}
\def\HiggsSignals{{\sc HiggsSignals}\xspace}
\def\HiggsBounds{{\sc HiggsBounds}\xspace}
\def\HiggsTools{{\sc HiggsTools}\xspace}
\def\Vevacious{{\sc Vevacious}\xspace}
\def\Vevaciouspp{{\sc Vevacious++}\xspace}

\def\MadGraph{{\sc MG5\_aMC}\xspace}
\def\HackAnalysis{{\sc HackAnalysis}\xspace}
\newcommand{\pythia}{{\sc Pythia~8}\xspace}
\newcommand{\madanalysis}{{\sc Mad\-A\-na\-ly\-sis~5}\xspace}

\def\MultiNest{{\sc MultiNest}\xspace}
\def\Diver{{\sc Diver}\xspace}

\def\GAMBIT{{\sc GAMBIT}\xspace}
\def\BSMArt{{\sc BSMArt}\xspace}

\newcommand{\cq}[1]{{\tt "#1"}}

\definecolor{maroon}{cmyk}{0, 0.87, 0.68, 0.32}
\definecolor{halfgray}{gray}{0.55}
\definecolor{slha_frame}{RGB}{207, 207, 207}
\definecolor{slha_bg}{RGB}{247, 247, 247}
\definecolor{slha_red}{RGB}{186, 33, 33}
\definecolor{slha_green}{RGB}{0, 128, 0}
\definecolor{slha_cyan}{RGB}{64, 128, 128}
\definecolor{slha_purple}{RGB}{170, 34, 255}

\definecolor{mathematica_frame}{RGB}{207, 207, 207}
\definecolor{mathematica_bg}{RGB}{247, 247, 247}
\definecolor{mathematica_red}{RGB}{186, 33, 33}
\definecolor{mathematica_green}{RGB}{0, 128, 0}
\definecolor{mathematica_cyan}{RGB}{64, 128, 128}
\definecolor{mathematica_purple}{RGB}{170, 34, 255}

\lstnewenvironment{MIN}[1][]{%
  \renewcommand{\thelstnumber}{In[\arabic{lstnumber}]}
  \lstset{language=MathIn,numbers=left,basicstyle=\ttfamily,#1}%
}{%
}

\lstnewenvironment{MOUT}[1][]{%
  \renewcommand{\thelstnumber}{Out[\arabic{lstnumber}]}
  \lstset{language=MathOut,numbers=left,basicstyle=\ttfamily,#1}%
}{%
}

\usepackage{listings}
\lstdefinelanguage{SLHA}{
    morekeywords={block,Block,BLOCK,decay,Decay,DECAY},%
    sensitive=true,%
    morecomment=[l]\#,%
    morestring=[b]',%
    morestring=[b]",%
    morestring=[s]{'''}{'''},% used for documentation text (mulitiline strings)
    morestring=[s]{"""}{"""},% added by Philipp Matthias Hahn
    morestring=[s]{r'}{'},% `raw' strings
    morestring=[s]{r"}{"},%
    morestring=[s]{r'''}{'''},%
    morestring=[s]{r"""}{"""},%
    morestring=[s]{u'}{'},% unicode strings
    morestring=[s]{u"}{"},%
    morestring=[s]{u'''}{'''},%
    morestring=[s]{u"""}{"""},%
    identifierstyle=\color{black}\ttfamily,
    commentstyle=\color{slha_cyan}\ttfamily,
    stringstyle=\color{slha_red}\ttfamily,
    keepspaces=true,
    showspaces=false,
    showstringspaces=false,
    rulecolor=\color{slha_frame},
    frame=single,
    frameround={t}{t}{t}{t},
    framexleftmargin=6mm,
    numbers=left,
    numberstyle=\tiny\color{halfgray},
    backgroundcolor=\color{slha_bg},
    basicstyle=\footnotesize,
    keywordstyle=\color{slha_green}\ttfamily,
    aboveskip=1.2em,
    belowskip=1.2em,
}

\lstdefinelanguage{MathIn}{
    morekeywords={Simplify,Eigenvalues},%
    emph={Start,InitUnitarity,GetScatteringDiagrams,BuildScatteringMatrix,MakeSPheno},%
    emphstyle={\color{mathematica_purple}},
    sensitive=true,%
    morecomment=[l]\%,%
    morestring=[b]',%
    morestring=[b]",%
    morestring=[s]{'''}{'''},% used for documentation text (mulitiline strings)
    morestring=[s]{"""}{"""},% added by Philipp Matthias Hahn
    morestring=[s]{r'}{'},% `raw' strings
    morestring=[s]{r"}{"},%
    morestring=[s]{r'''}{'''},%
    morestring=[s]{r"""}{"""},%
    morestring=[s]{u'}{'},% unicode strings
    morestring=[s]{u"}{"},%
    morestring=[s]{u'''}{'''},%
    morestring=[s]{u"""}{"""},%
    identifierstyle=\color{black}\ttfamily,
    commentstyle=\color{mathematica_cyan}\ttfamily,
    stringstyle=\color{mathematica_red}\ttfamily,
    keepspaces=true,
    showspaces=false,
    showstringspaces=false,
    rulecolor=\color{mathematica_frame},
    frame=single,
    frameround={t}{t}{t}{t},
    framexleftmargin=10mm,
    numbers=left,
    numberstyle=\tiny\color{halfgray},
    backgroundcolor=\color{mathematica_bg},
    basicstyle=\footnotesize,
    keywordstyle=\color{mathematica_green}\ttfamily,
    aboveskip=1.2em,
    belowskip=1.2em,
}

\lstdefinelanguage{MathOut}{
    morekeywords={Simplify,Eigenvalues},%
    sensitive=true,%
    morecomment=[l]\%,%
    morestring=[b]',%
    morestring=[b]",%
    morestring=[s]{'''}{'''},% used for documentation text (mulitiline strings)
    morestring=[s]{"""}{"""},% added by Philipp Matthias Hahn
    morestring=[s]{r'}{'},% `raw' strings
    morestring=[s]{r"}{"},%
    morestring=[s]{r'''}{'''},%
    morestring=[s]{r"""}{"""},%
    morestring=[s]{u'}{'},% unicode strings
    morestring=[s]{u"}{"},%
    morestring=[s]{u'''}{'''},%
    morestring=[s]{u"""}{"""},%
    identifierstyle=\color{black}\ttfamily,
    commentstyle=\color{mathematica_cyan}\ttfamily,
    stringstyle=\color{mathematica_red}\ttfamily,
    keepspaces=true,
    showspaces=false,
    showstringspaces=false,
    rulecolor=\color{mathematica_frame},
    frame=single,
    frameround={t}{t}{t}{t},
    framexleftmargin=10mm,
    numbers=left,
    numberstyle=\tiny\color{halfgray},
    backgroundcolor=\color{mathematica_bg},
    basicstyle=\footnotesize,
    keywordstyle=\color{mathematica_green}\ttfamily,
    aboveskip=1.2em,
    belowskip=1.2em,
}

\let\origthelstnumber\thelstnumber
\makeatletter
\newcommand*\Suppressnumber{%
  \lst@AddToHook{OnNewLine}{%
    \let\thelstnumber\relax%
     \advance\c@lstnumber-\@ne\relax%
    }%
}

\newcommand*\Reactivatenumber{%
  \lst@AddToHook{OnNewLine}{%
   \let\thelstnumber\origthelstnumber%
   \advance\c@lstnumber\@ne\relax}%
}

\numberwithin{equation}{section}

\title{BSMArt: Fast and efficient parameter space scans}

\date{}

\begin{document}

\begin{flushright}
\end{flushright}
\begin{center}

\vspace{1cm}
%{\bf \LARGE BSMArt: a suitable subtitle}
{\bf \LARGE BSMArt: simple and fast parameter space scans}

\vspace{1cm}

\large{Mark D. Goodsell\footnote{goodsell@lpthe.jussieu.fr} and
Ari Joury\footnote{ari@joury.eu}
 \\[5mm]}

{ \sl Sorbonne Universit\'e, CNRS, Laboratoire de Physique Th\'eorique et Hautes Energies (LPTHE), F-75005 Paris, France. }

\end{center}
\vspace{0.7cm}

\abstract{We introduce BSMArt, a python program for the exploration of parameter spaces of theories Beyond the Standard Model. Especially designed for use with the SARAH family of tools, it is also sufficiently flexible to be used with a wide variety of external codes. BSMArt contains the first public release of the Active Learning scan by the same authors; but contains several additional scanning algorithms, ranging from the very simple to MultiNest and Diver. A BSMArt scan can be set up in a matter of minutes with only minimal editing of configuration files; installation scripts for all relevant tools and examples are provided.
}

\newpage
\setcounter{footnote}{0}
\tableofcontents
%%%%%%%%%%%%%%%%%%%%%%%%%%%%%%%%%%%%%%%%%%%%%%%%%%%%%%%%%%%%%%%%%%%%%%%%%%%%%%%%%%%%%%%%%%%%%%%%%%%%%%%%%%%%%%%%%%%%%%%

\section{Introduction}
\label{SEC:INTRO}

The search for new physics at the energy frontier has been greatly facilitated in recent years by the development of a host of generic tools, designed to compare almost any model to the latest data. While this program of research is still in rapid development, there are by now a well-established set of codes available to phenomenologists looking to test their new theory, or explore new regions of an old one. Indeed, the need for generic codes, rather than those tailored just to specific models, has greatly grown since the days when the Minimal Supersymmetric Standard Model was almost the only example worth considering. 

Out of those supersymmetric days grew the SUSY Les Houches accords (SLHA) \cite{Allanach:2008qq}, a format for allowing codes to communicate with each other by exchanging text files. Howevever, there was no standard framework for automating the task of making the codes talk to each other, so for any new model a new set of programs would have to be written by hand. Hence collaborations such as {\tt MasterCode} \cite{Buchmueller:2007zk,Bagnaschi:2019djj} or {\tt BayesFITS} \cite{Darme:2019wpd} have a collection of private codes to do this (see also \cite{Ahmed:2022jlo}). To solve this problem generally, the code {\tt SARAH Scan and Plot (SSP)} \cite{Staub:2011dp} was written, that could chain the codes together using {\tt Mathematica} and enable simple scans. However, that code is no-longer supported, and suffers from many drawbacks.

More recently, \GAMBIT~\cite{GAMBIT:2017yxo,Kvellestad:2019vxm,Bloor:2021gtp} pioneered a different approach, circumventing the SLHA files by creating a ``backend'' for each tool. In this way, all of the programs are compiled into one code and the different components communicate internally without the need for text files. It was also equipped with a sophisticated suite of scanning tools. While it is highly recommended for high-powered exploration of model parameter spaces, and the only public tool that will allow a global confrontation of a model with data, it is not suitable for every use case: (1) It was designed with the aim of reconstructing posterior distributions of likelihood functions computed from the ensemble of tools involved; as described in \cite{Goodsell:2022beo}, for simple initial explorations or for many applications -- such as finding just exclusion boundaries where the decision is discrete (excluded/allowed) or where the user is only interested in specific observables -- this is not convenient. Sometimes a black box is not what is required. (2) It is not straightforward to change the codes involved -- in particular, the versions of codes employed are fixed at the moment the backend is created (or updated); and this makes using experimental versions of codes very difficult and means that the workflow is not straightforward to change. (3) It is daunting (even if, in fact, it is not complicated) to implement user-defined scans; but this is especially so to use Machine Learning, where many libraries exist in {\tt python} rather than {\tt c++}.

With this in mind, there is a need for a simpler, more flexible, general-purpose tool that can be run on a single computer, for exploratory rather than global studies. This is what \BSMArt is indended to be. It grew out of an initial collaboration to create {\tt xBit} \cite{Staub:2019xhl}, but is a complete rewrite in order to overcome that tool's own shortcomings and add different features. \BSMArt is thus a {\tt python} program that calls the required tools externally and manages the exchange and reading of SLHA files. One fundamental reason for a rewrite is that the part of the code responsible for managing the running of external programs can now operate as a standalone module without requiring the scanning machinery and main executable: it can just be used as a black box that gives outputs from input variables, which makes interfacing with other programs much more straightforward. Moreover, it is written in such a way that only python packages that are actually needed are imported. But overall, the aim of \BSMArt is to make the process from writing down a model to its exploration as simple, fast and flexible as possible, and so we include scripts to download and build all the relevant tools which can be used, and to create all the necessary files from \SARAH \cite{Staub:2013tta,Goodsell:2017pdq,Braathen:2017izn,Goodsell:2018tti,Goodsell:2020rfu,Benakli:2022gjn} and set them up with \BSMArt, effectively also updating the (now obsolete) BSM toolbox \cite{Staub:2011dp}.

Early versions of \BSMArt have already been used in the preparation of several papers \cite{Goodsell:2020rfu,Goodsell:2022beo,Bagnaschi:2022zvd,Benakli:2022gjn}. It greatly simplifies the task of examining the properties of new and old models, leveraging the multiple cores of modern machines, where a new scan can be set up very rapidly just by editing a few lines in a configuration file provided by the scripts and a handful of lines in a spectrum file provided by \SARAH. But, as the name implies, it was especially designed for artificial intelligence applications to BSM; in this respect it implements the Active Learning algorithm of \cite{Goodsell:2022beo}, which is now made public for use by the community.

%%%%%%%%%%%%%%%%%%%%%%%%%%%%%%%%%%%%%%%%%%%%%%%%%

\section{Installation and running \BSMArt}
\label{SEC:INSTALLATION}

\BSMArt is available from \url{https://goodsell.pages.in2p3.fr/bsmart/}. It can be downloaded and run as a standalone package, if the user has already installed the various tools that they want to use. However, since one of the aims of the package is to replace the entire BSM toolbox \cite{Staub:2011dp} (including {\tt SSP}) there are several scripts provided to facilitate the installation and setup of all necessary tools; and the subsequent generation of code for a specific model. These are all available as a tarball called {\tt HEPtool\_install} from  \url{https://goodsell.pages.in2p3.fr/bsmart/QuickStart/} and include:
\begin{itemize}
\item {\tt installHEPtools.py}, a script to install all or any of the desired tools.
\item {\tt QuickStart.py} and two template input files. This sets up two scans involving the MSSM: a full (and somewhat slow) toy Markov-Chain-Monte-Carlo scan ({\tt QuickStart\_MSSM.json}) involving all of the tools installed by {\tt installHEPtools.py} and a `lightning' one ({\tt lightning\_MSSM.json}) that runs just \SPheno \cite{Porod:2011nf} very quickly. 
\item {\tt prepareModel.py}, a script to handle the running of \SARAH, compilation of the output code, and creation of a \BSMArt run directory and templates for the input files, for any model. 
\end{itemize}
To start using \BSMArt in the least possible time, the user need only install the tools, create the QuickStart MSSM code and run it by typing four commands:
%\begin{table}[h]\centering
%\renewcommand\tablename{Code}
\begin{lstlisting}[language=MathIn,basicstyle=\scriptsize,numbers=none]
  ./installHEPtools.py --All
  ./QuickStart.py
  cd BSMArt_QuickStart
../BSMArt_v1.0/bin/BSMArt QuickStart_MSSM.json
\end{lstlisting}
% \end{table}
\BSMArt will only run on *nix based systems (linux, unix, MacOS, Windows subsystem for linux) because it relies on unix shell programs for certain parts of its operation. It has also only been extensively tested on linux machines with minimal testing on MacOS. On the other hand, it has been designed to use as few python packages as possible, only importing additional ones as required (for example {\tt pytorch} is only imported if the {\tt AL} scan is used) in order to be compatible with older machines. It has been tested using {\tt python 3.5} and more recent versions; this is the reason that only a very basic status bar is used, for example -- that way additional packages (such as {\tt rich} or {\tt curses}) are not required.

Further details are provided below.

\subsection{Using the install script}

To install \BSMArt and all necessary tools the user need only invoke:
%\begin{table}[h]
\renewcommand\tablename{Code}
\begin{lstlisting}[language=MathIn,basicstyle=\scriptsize,numbers=none]
   ./installHEPtools.py --All  
\end{lstlisting}
%\end{table}
This will take some time to download and compile. 
Alternatively individual codes from the list \SARAH, \SPheno, \BSMArt, \MicrOMEGAs, \HiggsBounds, \HiggsSignals, \HiggsTools, \Vevaciouspp, \MultiNest and \Diver can be installed by individual flags, e.g.
%\begin{table}[h]
\renewcommand\tablename{Code}
\begin{lstlisting}[language=MathIn,basicstyle=\scriptsize,numbers=none]
   ./installHEPtools.py --MultiNest
\end{lstlisting}
%\end{table}
The code can be run more than once, to install different tools sequentially. 
Some configuration options are possible, e.g. the fortran compiler to be used with \SPheno via the flag {\tt --Fortran=} or the {\tt mpi} fortran compiler for \Diver via {\tt FF=}. Additional options are:
\begin{itemize}
\item {\tt --Reinstall} to wipe any existing version found, without removing downloaded tarballs.
\item {\tt --Redownload} to wipe any existing version found, including removing downloaded tarballs.
\item {\tt --AddPaths} to add the path to \SARAH to the downloaded version in your {\tt Mathematica} kernel. By default this is not done.
\item {\tt --LocalUrls} to use only the {\tt urls} stored in the download script, rather than trying to update them from the internet. By default, the script looks for an updated list before running, to allow easy increment to the latest version. 
\end{itemize}

After running, the script will create a file {\tt HEPtoolpaths.json} containing the paths to all installed tools. This can be used to configure the running of \BSMArt. 
Note that there are additional tools related to collider physics that can be used by \BSMArt that are \emph{not} installed by the above script. Instead, the installation and configuration of these is handled by a different script called {\tt installColliderTools.py} available from the {\tt HackAnalysis} \cite{Goodsell:2021iwc} website: \\\url{https://goodsell.pages.in2p3.fr/hackanalysis/}.
If the user wants to use these tools, they should run {\tt installColliderTools.py} before passing to the next step, because it will also add the paths of those tools to {\tt HEPtoolpaths.json}.

\subsection{Preparing the model}

After the installation phase, {\tt HEPtoolpaths.json} will have been created, containing paths to all necessary tools, e.g.:
%\begin{table}[h]
%\renewcommand\tablename{Code}
\begin{lstlisting}[language=MathIn,title=HEPtoolpaths.json,basicstyle=\scriptsize,numbers=none]
{
  "SARAH": "/home/user/BSMArt/SARAH-4.15.1",
  "SPheno": "/home/user/BSMArt/SPheno-4.0.5",
  "HiggsBounds": "/home/user/BSMArt/HiggsBounds-5.10.2",
  "HiggsSignals": "/home/user/BSMArt/HiggsSignals-2.6.2",
  "HiggsTools": "/home/user/BSMArt/higgstools-1.0",
  "MicrOMEGAs": "/home/user/BSMArt/micromegas_5.3.35",
  "BSMArt":"/home/user/BSMArt/BSMArt-1.0"
}
\end{lstlisting}
%\end{table}

Provided this file is in the same directory or findable in your {\tt \$PATH}, they can then be used by {\tt prepareModel.py} to prime your model for use with \BSMArt, e.g. via: 
%\begin{table}[h]
%\renewcommand\tablename{Code}
\begin{lstlisting}[language=MathIn,basicstyle=\scriptsize,numbers=none]
   ./prepareModel.py --All <ModelName>
\end{lstlisting}
%\end{table}
This will always attempt to create \SPheno output via \SARAH for the model of choice and compile the produced code; the flag {\tt --All} specifies that all available tools should be set up too, so that output for \MicrOMEGAs and \Vevaciouspp (if installed) should also be generated and the necessary code compiled. In particular, \BSMArt uses by default a modified main program for \MicrOMEGAs compared to that provided in \SARAH, as described below. The first phase of running of the program is to write a {\tt Mathematica} script to call \SARAH; it is therefore possible to configure the choices using {\tt --Init=} for commands to be passed to \SARAH \emph{before} invoking {\tt MakeSPheno} and {\tt --Options=} for a string of options to pass to {\tt MakeSPheno}. Most users will not require these, but for example the command
\begin{lstlisting}[language=MathIn,basicstyle=\scriptsize,numbers=none]
   ./prepareModel.py MSSM --Options="SkipFlavorKit=True;" --Init="TwoLoop->False,IncludeLoopDecays->False"
\end{lstlisting}
passes the script
\begin{lstlisting}[language=MathIn,basicstyle=\scriptsize,numbers=none]
  << "<SARAH path>/SARAH.m";
  Start["MSSM"];
  SkipFlavorKit=True;
  MakeSPheno["TwoLoop->False,IncludeLoopDecays->False"];
\end{lstlisting}
to {\tt Mathematica}, which will create the \SPheno code for the MSSM while disabling the {\tt FlavorKit} \cite{Porod:2014xia}and disabling two-loop RGEs and loop decays, all of which can be slow to run -- and then the script will move the the generated code to the \SPheno directory and compile it.

After creating and compiling necessary code (as well as extracting information about the number of Higgs bosons in the model for use with \HiggsBounds) the script will create a directory {\tt BSMArt\_<ModelName>} and place in it:
\begin{itemize}
\item A template {\tt json} file for defining a \BSMArt scan.
\item The template {\tt Les Houches} input files for the \SPheno code for the model. 
\end{itemize}
The configuration of these files and running \BSMArt will be described below. However, it is not \emph{necessary} to run \SPheno as part of a scan -- for example, \BSMArt could be used with outputs generated from {\tt FeynRules} -- in which case the {\tt prepareModel.py} script would be unnecessary.

\subsection{Running BSMArt}

\BSMArt is invoked on the command line by:
\begin{lstlisting}[language=MathIn,numbers=none]
.<BSMArt path>/bin/BSMArt <input json file>
\end{lstlisting}
All the settings for the run are therefore given in the {\tt json} file. 
Optionally, the flag {\tt --debug} can be added to print detailed debugging information to the screen and place additional information in the log files. The log files are by default (on machines that support them) placed in {\tt /dev/shm/BSMART\_Temp} (this can be overridden by an the option \cq{Temporary\ Directory} in \cq{Setup}  of the input json file). If {\tt /dev/shm} is absent or the {\tt --debug} flag is invoked they are placed instead in the subdirectory {\tt Temp} of the run directory.

Certain scans based on external libraries (\MultiNest and \Diver) use {\tt MPI} to handle multi-core processing; to use this feature \BSMArt should be launched by the {\tt MPI} executable:
\begin{lstlisting}[language=MathIn,numbers=none]
mpiexec -n <cores> <BSMArt path>/bin/BSMArt <input json file>
\end{lstlisting}
or
\begin{lstlisting}[language=MathIn,numbers=none]
mpiexec -n <cores> python3 -m mpi4py <BSMArt path>/bin/BSMArt <input json file>
\end{lstlisting}
The use of {\tt MPI} with python requires installation of the package {\tt mpi4py}.

The input json file must contain inputs for {\tt "Codes"} and {\tt "Setup"}. Most scans will also contain {\tt "Variables"}; {\tt "Observables"} and {\tt "Plotting"} can also be specified. A simple example (as generated by the {\tt QuickStart.py} script)  would be 
\begin{lstlisting}[language=MathIn,title=lightning\_MSSM.json,numbers=none]
{
  "Codes" : {
    "SPheno":{
             "Command": "/home/username/HEPTools/SPheno-4.0.5/bin/SPhenoMSSM",
             "InputFile": "lightning.in.MSSM",
             "Run": "True",
             "Observables":
             {
               "mh" : { "SLHA": ["MASS", [25]]}
             }
           }
    },
  "Setup" : {
    "RunName": "Lightning",
    "Type": "Random",
    "csv": "True",
    "Cores": 1,
    "Merge Results": "True",
    "StoreEverything": "False",
    "StoreAllPoints": "True",
    "StoreSeparateFiles": "False",
    "Output File": "MSSM_Output",
    "Spectrum File": "SPheno.spc.MSSM",
    "Points":100
  },
  "Variables":{
    	"m0": { "RANGE": [100,3000.0]},
    	"m12": { "RANGE": [100,3000.0]}
      },
   "Plotting":{
     "Strategy": "csv",
     "Plots": {
       "m0_vs_mh": { "Labels": ["$m_0$ (GeV)","$m_{h}$ (GeV)"], "Values": ["m0","mh"], "Ranges":[[100,3000],[118,122]]},
       "m12_vs_mh": { "Labels": ["$m_{1/2}$ (GeV)","$m_{h}$ (GeV)"], "Values": ["m12","mh"], "Ranges":[[100,3000],[118,122]]}
              }
	}   
      
}
\end{lstlisting}
The information in the {\tt "Setup"} dictionary will depend on the type of scan, with the exception of certain universal settings; in the above example only {\tt "Points"} is not universal. The {\tt "Cores"} option allows the user to specify the number of cores to be used for multi-core running; for most scans {\tt python} {\tt multiprocessing} library is used to ennable this, but this is not compatible with \MultiNest or \Diver scans which must be run through {\tt mpiexec} (described below). Furthermore, a value greater than 1 is not compatible with running \MadGraph which handles its own multi-processing.

  The dictionary {\tt "Codes"} provides at a minimum information about which codes to run (if {\tt "Run":"True"} is set), and the path to the executable or library as required. The information required for the included tools is detailed in section \ref{SEC:TOOLS}.

\subsubsection{Variables}

\BSMArt uses {\tt python}'s dictionaries and ability to transform strings to expressions in order to make specifying variables and observables as simple as possible. Each variable used by the scan, defined in the {\tt "Variables"} block, is a dictionary with its name as the key; the other options (in the above case {\tt "RANGE"}) depend on the type of scan used; see section \ref{SEC:SCANS} for details of each. The name of the variable can then be used directly in the template input file and its value will be automatically substituted at runtime. For scans using \SPheno, an input file must be specified (with name {\tt "InputFile"} in the settings for \SPheno) and in the run directory from which the code is launched (or at least discoverable in the system path), and the variables to be scanned over can be directly specified in that file via their names, e.g.:
\begin{lstlisting}[language=MathIn,title=lightning.in.MSSM,numbers=none]
Block MODSEL      #  
 1 1               #  1/0: High/low scale input 
 2 1              # Boundary Condition  
 6 1               # Generation Mixing 
Block SMINPUTS    # Standard Model inputs 
 2 1.166370E-05    # G_F,Fermi constant 
 3 1.187000E-01    # alpha_s(MZ) SM MSbar 
 4 9.118870E+01    # Z-boson pole mass 
 5 4.180000E+00    # m_b(mb) SM MSbar 
 6 1.728900E+02    # m_top(pole) 
 7 1.776690E+00    # m_tau(pole) 
Block MINPAR      # Input parameters 
 1  m0    # m0
 2  m12    # m12
 3  1.0000000E+01    # TanBeta
 4  1.0000000E+00    # SignumMu
 5  -2.0000000E+03    # Azero
Block SPhenoInput   # SPheno specific input 
  1 -1              # error level 
  2  0              # SPA conventions 
  7  1              # Skip 2-loop Higgs corrections 
  8  3              # Method used for two-loop calculation 
  9  1              # Gaugeless limit used at two-loop 
 10  0              # safe-mode used at two-loop 
 11 0               # calculate branching ratios
\end{lstlisting}
In this (lightning) example the values for $m_0$ and $m_{1/2}$ in the CMSSM will be scanned over; the strings {\tt m0} and {\tt m12} will be substituted at run time. 

The expressions can even be formulas and include expressions from {\tt python}'s {\tt math} library; it also applies to scans that do not use \SPheno; for example ones using \MadGraph or {\tt HackAnalysis} might use an input {\tt SLHA} file with {\tt DECAY} blocks such as:
\begin{lstlisting}[language=MathIn,numbers=none]
  DECAY 1000024 1.97e-14/ctau
\end{lstlisting}
where {\tt ctau} is then defined as a variable for scanning.

Finally, the input {\tt SLHA} file or some parts of it can also be directly specified in the {\tt json} scan file via {\tt "Blocks"}, e.g.
\begin{lstlisting}[language=MathIn,numbers=none]
  {
    "Blocks" : {
      "MINPAR": {
        "1": "m0",
        "2": "m12"
      }

    },
    "Setup": ...
    }
\end{lstlisting}
If the input {\tt SLHA} is found, the above will supercede the corresponding blocks in that file (so this could be useful for changing one or two settings); alternatively, it is not even necessary to supply the input file if  \emph{all} of the inputs are provided in this way. 

\subsubsection{Observables and collecting results}
\label{sec:observables}

Scans can have many different aims; the goal may be to sample only in a global likelihood for each point, or to find whether given points are excluded by different observations, or to observe the variation of different observables in the model. \BSMArt is intended to be as flexible as possible to accommodate all of these cases. The output from each code is stored during runtime as much as possible in {\tt SLHA}-like format in a temporary file specified by {\tt "Spectrum\ File"} in {\tt "Setup"}; it is then possible to extract ``observables'' from this file or to just store the whole thing for processing after the scan (or, indeed, both). The extraction of observables can be very useful, because instead of storing the entire spectrum file (which is often large), only the important information can be stored (in the case of comma-separated-values output or tabbed output) or passed to other codes or scripts. Observables can either be read after the execution of the relevant tool -- and therefore are defined as one of the options in the definition of the code, as in the {\tt lightning\_MSSM.json} example above where the Higgs mass is read and denoted {\tt "mh"} after \SPheno execution, or an {\tt "Observables"} dictionary can be added to to the script, in which case the observables are only gathered after the execution of all tools. This is a matter of preference, but if defined as part of the tool options then the observable is safely omitted if the tool is not run.

Observables may have different properties that need to be specified, depending on the scan type; for example, for any scan involving a likelihood, they should specify a type of scaling, a mean and a variance, because only specified observables will be used to compute the likelihood. But the only required field for an observable is  {\tt "SLHA"}, which points to a two-item list, the two entries being the name and number of the block in the {\tt SLHA} file to which the observable corresponds (blocks with two are more dimensions are dealt with by e.g. {\tt "SLHA": ["YU",[1,1]]}).

The results of a scan are always stored in a directory given by the {\tt "RunName"} in the {\tt "Setup"} dictionary. While certain scans may store additional outputs (for example, \MultiNest scans store the output from that code in a the directory {\tt <RunName>/MultiNest}) the actual parameter points and their results can be stored in different ways in the directory {\tt <RunName>/Spectrum\_Files} according to options in {\tt "Setup"}:
\begin{itemize}
\item {\tt "StoreAllPoints"}: if set true, the entire spectrum file for each point will be stored.
\item {\tt "StoreSeparatePoints"}: if true, each spectrum file will be stored and individually numbered. If false, the files will be concatenated, with points separated by {\tt ENDOFPARAMETERPOINT}. In both cases, the name of the file is given by the {\tt "Output File"} setting; in the case of separate files, the point number is appended to the name.
\item {\tt "StoreEverything"}: if true, then \emph{all} inputs and outputs of all of the codes are stored in separate, numbered, directories. This is useful for codes that produce several files as outputs or for being able to rerun the codes on individual points.
\item {\tt "StoreInputs"}: if true, the generated  input files are stored in a separate directory and numbered.
\item {\tt "csv"}: if true, a comma-separated-values file (named as the {\tt "Output File"} but with {\tt .csv} appended) is created that stores only the number of the point, the variables, outputs and the result of any postprocessing. A header line appears at the beginning of the file with the names of all of the entries, with the names of variables and observables. In this way it can be easily read as a {\tt pandas} dataframe.
\item {\tt "Short"}: identical to {\tt "csv"} except the results are saved separated by spaces with extension {\tt .dat}.
\item {\tt "Store In Memory"}: if true, all generated points are stored in lists internally for the duration of the scan. This may be useful for scans that want to train a network on the whole dataset; it can also be used by the plotting routines (so that plots are generated without reading text files).
\end{itemize}
All of these options are assumed false by default so need not be specified.

Note that reading and writing of {\tt SLHA} files is performed by an internal script called {\tt zslha.py}, which is based on {\tt xslha} \cite{Staub:2018rih} \footnote{See also {\tt PySLHA} \cite{Buckley:2013jua} which has no relation with this code but was written substantially earlier; and also {\tt pylha}.} but has been extensively modified and extended. Hence, if the user wants to read in the full spectrum files (either as separate files or as one concatenated file) after a scan, either {\tt xslha} or {\tt zslha.py} can be used 

\subsection{Plotting}
\label{SEC:PLOTTING}

\BSMArt includes some rudimentary routines for plotting, to enable simple two-dimensional plots of data to be automatically created after a run. This requires {\tt matplotlib}. Different strategies are possible: either {\tt "Strategy": "csv"} where the values from a stored comma-separated-values file are read; or {\tt  "Strategy":"SLHA"} where the information is read from a concatenated {\tt SLHA} file; or, if another value is given for the strategy, it will attempt to use values stored in memory (this will not work for every kind of scan, depending on how the run points are stored). 
Other than the strategy, each plot is specified in the input {\tt json} as a dictionary with the following elements:
\begin{itemize}
\item The key becomes the filename for the plot. For {\tt SLHA} and {\tt csv} plots a {\tt python} script will be created that can reproduce the plot, as well as the {\tt pdf} of the plot itself, placed in {\tt <RunName>/Plots}. So in the {\tt lightning\_MSSM.json} above we will create {\tt m0\_vs\_mh.py, m0\_vs\_mh.pdf,m12\_vs\_mh.py,m12\_vs\_mh.pdf}.
\item {\tt "Values"} are the values from the data to be used. In the case of {\tt csv} or memory plots these are the names of variables or observables; in an {\tt SLHA} plot these must be given in an identical format to defining an Observable, as a list containing the block name and number.
\item {\tt "Labels"} are the \LaTeX\ labels for the plot axes.
\item {\tt "Ranges"} are optional ranges for the axes.
\item {\tt "Scaling"} can specify a list of two items of  {\tt "log"},  {\tt "linear"},  {\tt "symlog"} etc corresponding to the {\tt [xscale,yscale]}, e.g. {\tt "Scaling": ["linear","log"]}.
\end{itemize}
The results are rather basic scatter plots but they should provide a rapid visualisation and, by providing the scripts to generate them, a basis for paper-ready images.

\section{BSMArt tools}
\label{SEC:TOOLS}

\BSMArt is fundamentally a tool for handling and simplifying the input, running and output of a range of BSM tools. Those supported with the initial release include:
\begin{itemize}
\item \SPheno \cite{Porod:2003um,Porod:2011nf} code generated from \SARAH \cite{Staub:2008uz,Staub:2013tta,Goodsell:2015ira,Braathen:2017izn,Goodsell:2017pdq,Goodsell:2018tti,Goodsell:2020rfu,Benakli:2022gjn}
\item \MicrOMEGAs \cite{Belanger:2007zz,Belanger:2010pz,Belanger:2018ccd}
\item \HiggsBounds \cite{Bechtle:2008jh,Bechtle:2011sb,Bechtle:2013wla,Bechtle:2020pkv}
\item \HiggsSignals \cite{Bechtle:2013xfa,Bechtle:2020uwn}
\item \HiggsTools \cite{Bahl:2022igd}
\item \Vevaciouspp \cite{Camargo-Molina:2013qva}
\item \HackAnalysis \cite{Goodsell:2021iwc}
\item \MadGraph \cite{Alwall:2011uj}
\item {\tt flavio} \cite{Straub:2018kue}
\end{itemize}
It is straightforward to incorporate new codes because the handling is performed by a python script for each tool, and the management of the working directory is performed by \BSMArt; this is described below. The generation of the code tailored to a specific model is supposed to be performed by \SARAH and this can be automated using the {\tt prepareModel.py} script.

When a scan is set up, the list of tools in the {\tt json} input file are read and the appropriate handling script loaded, taking care of any required initialisation. Any code-specific inputs are also given (these will be described for each tool below). The only options universal to every tool are the option \cq{Run} (if this is set to \cq{True} then the code is loaded) and \cq{Observables}, which were described in section \ref{sec:observables}. During running, for each point, each tool is run in the order that they are specified in the {\tt json} file, either until all codes have been run, or an exception is reached. However, there is also an advanced option whereby \cq{VALID} can be specified for a given observable. If defined within a code, then if the value of the observable is found to be outside that range then the subsequent codes will not be run (which can save expensive computations on uninteresting points). For example:
\begin{lstlisting}[language=MathIn,title=lightning\_MSSM.json,numbers=none]
{
  "Codes" : {
    "SPheno":{
             "Command": "/home/username/HEPTools/SPheno-4.0.5/bin/SPhenoMSSM",
             "InputFile": "lightning.in.MSSM",
             "Run": "True",
             "Observables":
             {
               "mh" : { "SLHA": ["MASS", [25]], "VALID":[122,128]}
             }
           },    
    "MicrOmegas":{
             "Command": "/home/username/HEPTools/micromegas_5.3.35/SARAH_MSSM/MicrOmegas_v5.2_BSMArt",
             "InputFile": "SPheno.spc.MSSM",
             "LSP": [1000022],
             "Run": "True",
             "DD_Limits": "False",
             "Observables":
             {
               "Oh2" : { "SLHA": ["DARKMATTER", [1]]}
             }
           }     
    },
  ...
      
}
\end{lstlisting}
In this case, \SPheno would be run on points and if the mass of the Higgs boson is found to lie within the range $[122,128]$ GeV then \MicrOMEGAs is executed (note the different capitalisation used within \BSMArt). Otherwise the point is labelled as not valid and no observables are returned for that point.

\subsection{\SPheno/\SARAH}

\BSMArt is designed especially for scans based on using \SPheno code from \SARAH as the first code. The running of this tool is especially simple; in addition to \cq{Run}, there are only two required settings:
\begin{itemize}
\item \cq{Command}: the \SPheno executable for your model
\item \cq{InputFile}: the {\tt SLHA} input file for \SPheno. As described above, the user may either provide a template file in the run directory, from which all other input files will be generated, or they must provide the inputs in the form of \cq{Blocks} in the {\tt json} input file -- or both. There are also some convenience functions for setting certain {\tt SLHA} inputs that can be called by a scan; this is advanced usage and can be found in {\tt core.py} in the {\tt bsmart} directory.
\end{itemize}
It is not necessary to specify an output filename for \SPheno as the \cq{Spectrum\ File} for the scan will be used (which is set to \cq{SLHA\_output} by default).

\subsection{\MicrOMEGAs}

\BSMArt is provided with two versions of a main program for use with \MicrOMEGAs, located in the {\tt AuxFiles} subdirectory. These are {\tt MicrOmegas\_v5.2\_BSMArt.cpp}, for use with recent versions, and {\tt MicrOmegas\_v5.0\_BSMArt.cpp} for use with versions prior to {\tt 5.2}, when a direct detection p-value was introduced. The {\tt prepareModel.py} script will automatically invoke \MicrOMEGAs {\tt newProject} command to create a direcory called {\tt SARAH\_<ModelName>}, copy the main program to it and the {\tt CalcHEP} files for the model, and build the main program.

The options for running \MicrOMEGAs within \BSMArt are:
\begin{itemize}
\item \cq{Command}: this should be the path to the {\tt MicrOmegas\_v5.2\_BSMArt} executable within the model directory in \MicrOMEGAs.
\item \cq{InputFile}: models built from \SARAH expect to find a spectrum file of the form {\tt SPheno.spc.<ModelName>} in the run directory; this name for a given {\tt <ModelName>} \emph{must} be specified as the \cq{InputFile} option.
\item \cq{OutputFile}: the {\tt MicrOmegas\_v5.2\_BSMArt} executable creates an output file called {\tt omg.out} by default and this will be assumed to be the case if no \cq{OutputFile} is specified; if the user modifies the main executable for \MicrOMEGAs and changes the output file then this can be specified.
\item \cq{LSP}: a list of allowed dark matter candidates for the model can be specified; if this option is given, then the code will check whether the lightest stable particle for each parameter point is in the list, and reject the point, halting code execution for it (either before or after running \MicrOMEGAs, depending on whether the spectrum file contains an {\tt LSP} block).
\item \cq{DD\_Limits}: earlier versions of \MicrOMEGAs did not compute a direct detection p-value, just providing the cross-sections, so \BSMArt contains a simple check against limits scraped from experimental plots. 
\end{itemize}
During execution, {\tt omg.out} is then appended to the spectrum file as a {\tt DARKMATTER} block, which might look like:
\begin{lstlisting}[language=MathIn,numbers=none]
Block DARKMATTER
1 1.674264e+01 # Total relic density Omega h^2 
2 1000022 # CDM1 
3 5.355477e+02 # CDM1 Mass (GeV) 
100 0.282639 # ~N1 ~N1 -> e3 E3
101 0.266913 # ~N1 ~N1 -> e2 E2
102 0.266860 # ~N1 ~N1 -> e1 E1
103 0.063635 # ~N1 ~N1 -> u3 U3
104 0.022296 # ~N1 ~N1 -> Wm Wp
105 0.016841 # ~N1 ~N1 -> u2 U2
106 0.016840 # ~N1 ~N1 -> u1 U1
107 0.012315 # ~N1 ~N1 -> nu3 Nu3
108 0.012021 # ~N1 ~N1 -> nu2 Nu2
109 0.012020 # ~N1 ~N1 -> nu1 Nu1
201 1.210892e-11 # 
202 8.079591e-09 #
203 1.247376e-11 #
204 1.175055e-08 # 
401 0.500000 # Direct detection pval
\end{lstlisting}
Entries between 100 and 199 are thus the relative contribution of different annihilation processes. 
Entries 201 to 204 are the CDM-nucleon cross sections in pb, as proton spin-independent, proton spin-dependent, neutron spin independent and neutron spin dependent respectively. If a second stable dark matter particle is present, entries 4,5,6 and 7 contain respectively the CDM1 relic density (as opposed to the total); the CDM2 pdg id number; the CDM2 mass and the CDM2 relic density.

\subsection{{\tt flavio}}

It is straightforward to run {\tt flavio} \cite{Straub:2018kue} as it is a {\tt python} module that can be imported. The running in \BSMArt is, however, rudimentary, in that the only option is toe specify an \cq{InputFile}, which must be a {\tt WCXF} file \cite{Aebischer:2017ugx}. The observables that can then be collected correspond to named observables within {\tt flavio}, which will be placed in order in {\tt BLOCK FLAVIO} in the spectrum file and/or used by \BSMArt internally. 

\subsection{\HiggsBounds and \HiggsSignals}

To run \HiggsBounds and \HiggsSignals, the relevant code entries might look like:
\begin{lstlisting}[language=MathIn,numbers=none]
"Codes":{
  "HiggsBounds":{
             "Command": "/home/username/BSMArt/higgsbounds-5.10.2/build/HiggsBounds",
             "Neutral Higgs": 3,
             "Charged Higgs": 1,
             "Run": "True"
           },
  "HiggsSignals":{
             "Command": "/home/username/BSMArt/higgssignals-2.6.2/build/HiggsSignals",
             "Neutral Higgs": 3,
             "Charged Higgs": 1,
             "Run": "True"
           }
}
\end{lstlisting}
Alternatively, instead of supplying the number of neutral and charged Higgs fields (which are extracted automatically by the {\tt prepareModel.py} script), the options to pass to the executable can be supplied by \cq{Options}, e.g.
\begin{lstlisting}[language=MathIn,numbers=none]
"Codes":{
  "HiggsBounds":{
             "Command": "/home/username/BSMArt/higgsbounds-5.10.2/build/HiggsBounds",
             "Options": "LandH effC 3 1",
             "Run": "True"
           },
  "HiggsSignals":{
             "Command": "/home/username/BSMArt/higgssignals-2.6.2/build/HiggsSignals",
             "Options": "latestresults 2 effC 3 1",
             "Run": "True"
           }
}
\end{lstlisting}
The only other possible setting is \cq{OutputFile} for each tool, which is otherwise assumed to be \cq{HiggsBounds\_results.dat} and \cq{HiggsSignals\_results.dat} for both tools respectively.

The output stored in the \HiggsBounds and \HiggsSignals files is tranformed into standard {\tt SLHA} format in blocks {\tt HIGGSBOUNDS} and {\tt HIGGSSIGNALS} in the spectrum files, in the form:
\begin{lstlisting}[language=MathIn,numbers=none]
BLOCK HIGGSBOUNDS #
  1  1.20776000E+02 #  Mh(1)
  2  2.29044000E+03 #  Mh(2)
  3  2.29041000E+03 #  Mh(3)
  4  2.29162000E+03 #  Mhplus(1)
  5  0.00000000E+00 #  HBresult
  6  3.86000000E+02 #  chan
  7  1.10804000E+00 #  obsratio
  8  1.00000000E+00 #  ncomb
101 1.10804000E+00 # HiggsBounds obs ratio
BLOCK HIGGSSIGNALS #
  1  1.20776000E+02 #  Mh(1)
  2  2.29044000E+03 #  Mh(2)
  3  2.29041000E+03 #  Mh(3)
  4  2.29162000E+03 #  Mhplus(1)
  5  9.67329000E+01 #  csq(mu)
  6  7.42778000E-01 #  csq(mh)
  7  9.74757000E+01 #  csq(tot)
  8  1.10000000E+02 #  nobs(mu)
  9  1.00000000E+00 #  nobs(mh)
 10  1.11000000E+02 #  nobs(tot)
 11  8.16521000E-01 #  Pvalue
101 8.16521000e-01 # HiggsSignals P value
\end{lstlisting}
\BSMArt automatically places the \HiggsBounds ratio of signal to limit in entry 101 (the point is excluded if this is greater than 1) and the \HiggsSignals p-value in entry 101 of the {\tt HIGGSSIGNALS}  block (again, the point is excluded if this is greater than one). The other entries are given in the order appearing in a single line in the output files from these tools; the user should consult the user manuals.

\subsection{{\tt HiggsTools}}

{\tt HiggsTools} \cite{Bahl:2022igd} is the recently-released rewrite of {\tt HiggsBounds} and {\tt HiggsSignals} that incorporates both into one code, has a more complete database and also includes a {\tt python} interface. In \BSMArt we make use of this interface, which saves the program certain overheads because the loading of databases of analyses and intialisation need only be performed once, rather than for each point. The passing of values to {\tt HiggsTools} is also simpler and does not require the creation of datafiles; all the information is taken from the {\tt slha} file, provided that the {\tt HIGGSCOUPLINGSBOSONS} and {\tt HIGGSCOUPLINGSFERMIONS} blocks are specified. This means, in an {\tt slha} input from \SARAH, the {\tt SPhenoInput} block should contain:
\begin{lstlisting}[language=MathIn,numbers=none]
BLOCK SPhenoInput
  11 1               # calculate branching ratios
 520 1.              # Write effective Higgs couplings (HiggsBounds blocks): put 0 to use file with MadGraph! 
\end{lstlisting}

It is also necessary to give the {\tt pdg} numbers for the neutral and charged Higgs bosons present in the model as lists via \cq{Neutral Higgs} and \cq{Charged Higgs} options. These can be automatically extracted from the model by \SARAH and written into the template by the {\tt prepareModel.py} script. So, for example, the {\tt QuickStart\_MSSM.json} file may contain: 
\begin{lstlisting}[language=MathIn,numbers=none,basicstyle=\scriptsize]
"Codes":{
  "HiggsTools":{
             "HiggsBounds Dataset": "/home/username/BSMArt/hbdataset-master",
             "HiggsSignals Dataset": "/home/username/BSMArt/hsdataset-main",
             "Neutral Higgs": [25, 35, 36],
             "Charged Higgs": [37],
             "Observables":{
                            "HBresult" : {"SLHA": ["HIGGSTOOLS",[1]],"SCALING":"OFF"},
                            "HBobsr" : {"SLHA": ["HIGGSTOOLS",[2]], "SCALING":"UPPER","MEAN":1.0,"VARIANCE":0.1},
                            "HSchi2" : {"SLHA": ["HIGGSTOOLS",[11]],"SCALING":"OFF"},
                            "HSnobs" : {"SLHA": ["HIGGSTOOLS",[12]],"SCALING":"OFF"},
                            "HSpval" : {"SLHA": ["HIGGSTOOLS",[13]], "SCALING":"LOWER","MEAN":0.001,"VARIANCE":0.1}
                           },

             "Run": "True"
           }
}
\end{lstlisting}
The results are processed internally and also written to the spectrum files in a {\tt HIGGSTOOLS} block. The important entries are given above as:
\begin{itemize}
\item {\tt 1}: the result of \HiggsBounds ({\tt 0} for excluded and {\tt 1} for allowed)
\item {\tt 2}: the maximum ratio of allowed to observed signal in \HiggsBounds.
\item {\tt 11}: the $\chi^2$ computed by \HiggsSignals
\item {\tt 12}: the number of observables included in the $\chi^2$ computation by \HiggsSignals
\item {\tt 13}: the $p$-value inferred by \BSMArt from the $\chi^2$ of \HiggsSignals (computed for backwards compatibility and comparison with \HiggsSignals)
\item Other entries given by the {\tt pdg} number of any of the Higgs bosons in analyses selected by \HiggsBounds  (i.e. those that give the strongest limit) with the result begin the observed ratio; the comments contain information about which analysis it is.
\end{itemize}
Hence for the MSSM a result might look like:
\begin{lstlisting}[language=MathIn,numbers=none]
BLOCK HIGGSTOOLS 
1  1  # HB result (0/1)
2  7.59226725e-01  # HB maximum obs ratio
11  1.25077926e+02  # HS chi2
12  131  # HS number of observables
13  6.29364754e-01  # BSMArt Inferred HS pval
25  7.59226725e-01  # ID: 12045, ref: CMS-PAS-HIG-12-045: LHC8 [vbfH,HW,Htt,H,HZ]>[bb,tautau,WW,ZZ,gamgam]
\end{lstlisting}

It is in principle possible to obtain a (currently experimental) computation of the mass uncertainty for the Higgs bosons from \SARAH/\SPheno code. If activated, this will be read and used by \HiggsTools; if absent, a mass uncertainty of 3 GeV will be assigned internally; otherwise \HiggsSignals will be overly pedantic about the SM-like Higgs boson and use only the (tiny) experimental uncertainty on its mass. 

It should be noted that \HiggsTools requires relatively recent versions of {\tt python} (so is not compatible with version 3.5); hence it may be necessary to use the original \HiggsBounds and \HiggsSignals interfaces on older machines.

\subsection{\MadGraph}

\BSMArt can run \MadGraph  on a series of generated parameter points. Running \MadGraph on its own is rather rudimentary and intended just for cross-section computations; more advanced usage is included in the {\tt MadGraphHackAnalysis} tool. As such, the only setting is \cq{Command} which should point to the {\tt generate\_events} executable within a directory created by \MadGraph for your process; the outputs are stored in the spectrum file in the form:
\begin{lstlisting}[language=MathIn,numbers=none]
Block MADGRAPH
1 0.0000000E+00 # Cross-section (pb)
2 0.0000000E+00 # + Fractional uncertainty
3 0.0000000E+00 # - Fractional uncertainty 
\end{lstlisting}
Hence these can be used as observables.

It should be noted that the output files from \SARAH/SPheno, while separately compatible with \MadGraph and \HiggsTools, will not work at the same time due to the nature of {\tt SLHA} blocks and the fact that we use the {\tt SLHA} interface for \HiggsTools. This will be remedied in a future version at the same time as an upgrade to \SARAH.

\subsection{{\tt HackAnalysis\_LO} and  {\tt MadGraphHackAnalysis}}

\subsubsection{Inputs}

To run recasting using the (limited number of) analyses available in {\tt HackAnalysis}, \BSMArt can run either using \pythia to generate events and shower directly, or \MadGraph to generate the events before showing by \pythia. For the former, {\tt analysePYTHIA.exe} is used, and the path to this executable should be given using the \cq{Command} option; running should be handled by the {\tt HackAnalysis\_LO} tool. For \MadGraph-generated events, the showering can either be performed by \MadGraph and the resulting {\tt hepmc} file read by {\tt analyseHEPMC.exe}; or the showering using \pythia can be performed using {\tt analysePYTHIA\_LHE.exe}. For both of these, the relevant tool is {\tt MadGraphHackAnalysis}. 

The relevant settings might look like:
\begin{lstlisting}[language=MathIn,numbers=none]
"Codes":{
  "HackAnalysis_LO":{
             "Command": "/home/username/BSMArt/hackanalysis-1.2/analysePYTHIA.exe",
             "YAML file": "LOpythia.yaml",
             "Run": "True"
           },
  "MadGraphHackAnalysis":{
             "MadGraph": "/home/username/BSMArt/MGMODELDIR/bin/generate_events",
             "HackAnalysis": "/home/username/BSMArt/hackanalysis-1.2/analysePYTHIA_LHE.exe",
             "YAML file": "MLM.yaml",
             "Run": "True"
           },
  "MadGraphHackAnalysis":{
             "MadGraph": "/home/username/BSMArt/MGMODELDIR/bin/generate_events",
             "HackAnalysis": "/home/username/BSMArt/hackanalysis-1.2/analyseHEPMC.exe",
             "YAML file": "HEPMC.yaml",
             "Run": "True"
           }         
}
\end{lstlisting}
The relevant information for running is read from the {\tt yaml} file, including information for \pythia (if applicable) and the number of cores to use. While the number of cores for the scan overall (in \cq{Setup}) should be set to 1 when using the collider tools (because they handle multicore operation themselves), the number of cores to use in \MadGraph is handled by the settings in the model directory; and the number of cores to use in \HackAnalysis \pythia runs is set as an option within the {\tt yaml} file (see the \HackAnalysis website for details). In particular, when generating events using \MadGraph and showering them within \HackAnalysis, \BSMArt will handle the splitting of the {\tt lhe} files.

For advanced usage, it may be desireable to pass additional options to \MadGraph and/or \pythia using model-dependent information. This is possible with an \cq{Extra\ Commands} option, which should be a list of commands to write to a text file that can then be passed to \MadGraph at runtime. If a command is itself a list, it is assumed that it is an expression to be evaluated, given in terms of the names of variables or observables of the point. For example, we might put:
\begin{lstlisting}[language=MathIn,numbers=none]
"Codes":{
  "MadGraphHackAnalysis":{
             "MadGraph": "/home/username/BSMArt/MGMODELDIR/bin/generate_events",
             "HackAnalysis": "/home/username/BSMArt/hackanalysis-1.2/analysePYTHIA_LHE.exe",
             "YAML file": "MLM.yaml",
             "Extra Commands": [['shower off'], ['0'],['set decay 1000024','1.97e-15/ctau'],['set xqcut', 'mCha/4']]
             "Run": "True"
           }  
}
\end{lstlisting}
Similarly, variables such as {\tt qcut} can be defined in the \pythia configuration file, specified in the {\tt yaml} file, and this can depend on variables/observables too. The template configuration file should be found in the run directory; at run time is it regenerated by the code. It should consist of a set of commands of the form `command = value':
\begin{lstlisting}[language=MathIn,title=MLM.cfg,numbers=none]
Init:showChangedSettings=off
Init:showChangedParticleData = off  ! not useful info here
Next:numberShowInfo       = 0    ! avoid unnecessary info
Next:numberShowEvent      = 0    ! avoid lengthy event listings
Next:numberCount=0

Beams:setProductionScalesFromLHEF=on

JetMatching:merge            = on
JetMatching:scheme           = 1
JetMatching:setMad           = off
JetMatching:qCut             = mCha/4
\end{lstlisting}
Whenever a string is found, \BSMArt will attempt to evaluate it as an expression against the variables/observables it has collected. So in this case, if the model has an observable called `{\tt mCha}' then the {\tt JetMatching:qCut} will be set for that point with the appropriate value. When using this tool, no variables/observables should be named `on' or `off'!

\subsubsection{Results}

For the purpose of \BSMArt runs, the results are the efficiency files, with a specified in the {\tt yaml} file; these will be appended to the spectrum file and any entries can be used as an observable. They have the form:
\begin{lstlisting}[language=MathIn,numbers=none]
BLOCK PROCESS ATLAS_SUSY_2017_04_3body
XS   0.690957   # (fb) 
Events  20003
BLOCK EFFICIENCIES ATLAS_SUSY_2017_04_3body # name, eff, rel. uncert.
SR   3.749438e-03   1.152534e-01
SR_ee   1.999700e-03   1.579557e-01
SR_emu   1.699745e-03   1.713528e-01
SR_mumu   0.000000e+00   0.000000e+00
\end{lstlisting}

\subsection{\Vevaciouspp}

\Vevacious \cite{Camargo-Molina:2013qva} was designed to run with \SARAH and \SPheno; however, it relies on codes that no longer run on modern machines, so only \Vevaciouspp is supported by \BSMArt. This tool requires:
\begin{itemize}
\item \cq{Command}: the path to the \Vevaciouspp executable
\item \cq{Initialization\ File}
\end{itemize}
There is also the possibility to specify \cq{InputFile} which is otherwise assumed to be \cq{VevaciousInput.xml}. The output from the code is appended to the spectrum file by \Vevaciouspp in {\tt SLHA} format, so can be readily used as observables by \BSMArt. 

\subsection{Adding a new tool}

To add a new tool called ``{\tt toolName}'', the user just needs to add a script with the name ``{\tt toolName.py}'' to the subdirectory {\tt bsmart/tools} of the installation. A template for this would be:
\begin{lstlisting}[language=MathIn,title=toolName.py,numbers=none]

import debug
import zslha
from HEPRun import  HepTool, DataPoint

class NewTool(HepTool):
    def __init__(self, name, settings,global_settings=None):
        HepTool.__init__(self, name, settings,global_settings)
        
        ## Initialisation here
        
    def run(self, spc_file, temp_dir, log,data_point):
        ## First run your tool

        ## Then place the outputs into spc_file and data_point.spc
        
\end{lstlisting}
Initialisation can thus be handled by overloading the {\tt \_\_Init\_\_(self, name, settings,global\_settings=None)} method, where {\tt settings} is the ordered dictionary specified in the relevant code name in the input {\tt json} file; and {\tt global\_settings} is an ordered dictionary corresponding to all of the run settings. The running is specified by the {\tt run(self, spc\_file, temp\_dir, log,data\_point)} function, which is handed the spectrum file, location of the temporary directory (where the code is actually run from), log (for error handling) and a {\tt DataPoint} object (defined in {\tt HEPRun.py}).

Users are strongly encouraged to submit any new tool interfaces that they create for inclusions in future releases!

\section{BSMArt scans}
\label{SEC:SCANS}

\BSMArt is designed to make running simple (or more complicated) scans as simple as possible, but also to allow maximum flexibility. This means that different work flows are supported, and it is very straightforward to implement new ad-hoc scans. A new scan is invoked via the {\tt Type} option of {\tt Setup} in the {\tt json} file:
\begin{lstlisting}[language=MathIn,numbers=none]
  "Setup" : {
    "Type" : "<scanname>",
    ...
  \end{lstlisting}
The initial release contains the following scan types:
\begin{itemize}
\item Grid.
\item Random.
\item MCMC.
\item {\tt read\_csv}.
\item {\tt read\_dir}.
\item AL \cite{Goodsell:2022beo}.
\item \MultiNest \cite{Feroz:2008xx,Feroz:2013hea} with inspiration from {\tt pyMultiNest} \cite{Buchner:2014nha}.
\item \Diver \cite{Martinez:2017lzg}.
\end{itemize}
We describe each of these in the following; examples are included with the code in the {\tt InputExamples} directory. \BSMArt is also designed to make implementing a new scan as straightforward as possible, where running the chain of tools can be treated as a black box taking a list of input values and yielding outputs in the form of a list of observables, or even performing some customisable post-processing. The creation of new scans is define in sec. \ref{sec:NewScans}.

\subsection{Grid and Random scans}

The simplest types of scan involve generating points either on a hypergrid, or randomly sampling from defined ranges. For a {\tt grid} scan, the distribution of points is specified by {\tt numpy} functions, e.g.: 
%\begin{table}[h]
%\renewcommand\tablename{Code}
\begin{lstlisting}[language=MathIn,title=SMSQQ\_Grid.json,basicstyle=\scriptsize]
  "Setup": {
    "RunName": "SMSQQ_Grid_01",
    "Type": "Grid",
    "StoreAllPoints":"False",
    "Cores": 10
  },
    "Variables": {
      "kappa": "np.geomspace(1.0e2, 1.0e14, num=10)",
      "ms2": "np.geomspace(1.0e6, 1.0e12, num=10)",
      "mqm2": "np.geomspace(1.0e6, 1.0e12, num=10)",
      "mqp2": "np.geomspace(1.0e6, 1.0e12, num=10)",
      "lambda": "np.linspace(2.8, 3.2, num=10)"
  },
\end{lstlisting}
%\end{table}
Here, 10 points are scanned in every dimension, thus one generates $10^5$ points in total. %
Because the total number of points is implicitly contained in the listing of variables, it is not necessary to specify it in the \texttt{Setup} block of the file as well.

With models where the dimensionality is large, the computing time of grid scans can quickly get out of hand because the desired number of points grows exponentially with the number of dimensions. %
Random scans do not offer much more help with such situations, either; they are specified similarly: %
%\begin{table}[h]
%\renewcommand\tablename{Code}
\begin{lstlisting}[language=MathIn,title=SMSQQ\_Random.json,basicstyle=\scriptsize]
  "Setup": {
  	 "RunName": "SMSQQ_Random_01",
  	 "Type": "Random",
  	 "csv":"True",
  	 "StoreAllPoints":"True",
  	 "Cores": 20,
  	 "Merge Results": "True",
  	 "Points": 50000
    },
     "Variables": {
        "kappa": { "RANGE": [1.5e5,1.85e5]},
        "ms2-mqp2":    { "RANGE": [1.0e8,1.5e9]},
        "mqm2-ms2":   { "RANGE": [1e8,5.0e9]},
        "mqp2":   { "RANGE": [2e8,1.8e9]},
        "lams":  { "RANGE": [2.8,3.2]}
    },
\end{lstlisting}
%\end{table}
Here we only need the ranges to sample from and the total number of points.

Both of these scans generate all the points initially in memory and then run them as a batch, so that multiprocessing can be used by setting the option of \cq{Cores} in \cq{Setup} to a value greater than 1.

\subsection{{\tt read\_dir} and {\tt read\_csv}}

For many purposes it may be useful to run a scan using either points defined externally, or pre-generated files. The {\tt read\_dir} and {\tt read\_csv} scans handle these cases. In the former, the only parameter to specify is an \cq{Input\_Dir}; it is assumed that every file in that directory is an input file, which should be passed to the chain of tools. There are thus no variables. For the latter, \cq{Input\_CSV\_File} must be specified, giving the full path to a comma-separated-values file (which could be the output of a previous scan ...); the first line should be a list of names for the columns (so that it can be read by {\tt pandas}). The variables specified in the scan will therefore take values according to those in the file.

Both scans are compatible with multiprocessing and can be triggered by setting the option of \cq{Cores} in \cq{Setup} to a value greater than 1.

%---------------------------------------------------------------------------------------------------------

\subsection{MCMC}

\BSMArt contains an implementation of the Metropolis-Hastings algorithm to give a toy Markov-Chain Monte Carlo scan. Multi-core running is made possible by starting a different chain on each core; the scan is stopped after a given number of points have been sampled (\cq{Points} in \cq{Setup}) or valid points found (\cq{Valid\ Points}) in \cq{Setup}). In this context, `Valid Points' refers to those kept by the chain, rather than just those that give a valid set of observables. Results of each chain are stored separately during running, but may be merged to a single file upon completion if \cq{Merge\ Results} is set to \cq{True} in \cq{Setup}.

The likelihood function sampled by this scan is created from the observables; each observable should have a label of \cq{SCALING}, and  may also require \cq{MEAN} and \cq{VARIANCE}. Denoting the mean as $m$, the variance as $\sigma$ and the value of the observable as $x$, the scaling function has the following values:
\begin{itemize}
\item \cq{OFF} if it should be ignored for the likelihood function.
\item \cq{LOG} denotes $1/(1+(x-m)^2/(2 \sigma^2))$.
\item \cq{UPPER} denotes a sigmoid function $1/(1+ \exp((x-m)/\sigma))$.
\item \cq{LOWER} denotes a sigmoid function $1/(1+ \exp(-(x-m)/\sigma))$.
\item \cq{BIAS} denotes a power function $\left(\frac{x}{m}\right)^\sigma$.
\item \cq{USER} means that the value of the observable should be used as a likelihood (i.e. for codes that compute a likelihood as an output).
\item \cq{LOGUSER} and \cq{EXPUSER} mean the logarithm or exponential of the observable.
\item For any other setting (e.g. \cq{GAUSS} or not specified) it is assumed that a standard Gaussian likelihood is used $\exp[ -(x-m)^2/(2 \sigma^2)]$
\end{itemize}
The total likelihood is given as the product of all the individual likelihoods. In this way the user has rather fine control over the scan, but there is no information about correlations.

Similarly, the \cq{Variables} must also have a \cq{RANGE} and \cq{VARIANCE} specified for each one, to determine valid values and average jump sizes.

The toy MCMC is a very robust scan that should be useful for preliminary explorations of model spaces using a handful of variables, especially because of the presence of the unconventional features in the likelihood function that we have introduced. Indeed, this scan (and early versions of \BSMArt) has already been used in several publications \cite{Goodsell:2020rfu,Benakli:2022gjn}. On the other hand, it (currently) lacks features such as a convergence criterion, or features such as maximum lengths for the Markov Chains~\cite{Brivio:2019ius}, that can make more sophisticated approaches attractive.

\subsection{\MultiNest and \Diver}

As mentioned above, several libraries exist implementing efficient scans based on exploring the likelihood function. These are large codes that must be compiled and then manage the selection of points themselves. However, interfacing them to a selection of tools is not trivial and can be a tedious task; \BSMArt can therefore handle all of this, leaving the user to focus on the physics results. Included in the initial release are interfaces to \MultiNest and \Diver. Both of these libraries use {\tt MPI} to handle multi-core operation, so \BSMArt should be run in single core mode (\cq{Cores}:1 in \cq{Setup}) and executed via {\tt mpiexec}, preferentially via:
\begin{lstlisting}[language=MathIn,basicstyle=\scriptsize]
  mpiexec -n <cores> python3 -m mpi4py BSMArt <input json>
\end{lstlisting}
Both these tools use a log-likelihood function, so the options are:
\begin{itemize}
\item \cq{OFF} if it should be ignored for the likelihood function.
\item \cq{UPPER} denotes a log-sigmoid function $-\log (1+ \exp((x-m)/\sigma))$.
\item \cq{LOWER} denotes a log-sigmoid function $-\log (1+ \exp(-(x-m)/\sigma))$.
\item \cq{BIAS} denotes the logarithm of a power function $\sigma \log \frac{x}{m}$.
\item \cq{USER} means that the value of the observable should be used as a log-likelihood (e.g. the $\chi^2$ value from \HiggsSignals).
\item \cq{LOGUSER} means the logarithm of the observable.
\item For any other setting (e.g. \cq{GAUSS} or not specified) it is assumed that a standard $\chi^2$ is used $-(x-m)^2/(2 \sigma^2).$
\end{itemize}
The total likelihood is given as the \emph{sum} of all the individual log-likelihoods; {\tt Diver} requires this to also be multiplied by $-1$ which is done in postprocessing. 

Both scans require the variables to be specified with a \cq{RANGE}.

\MultiNest requires \cq{Live\ Points} or \cq{Points} to be specified in \cq{Setup}; and a path to the library specified by the option \cq{MultiNestLib} (if it is not in the environment path). Other options include \cq{Sampling Efficiency} and \cq{Evidence Tolerance}. For the meaning of these see \cite{Feroz:2008xx,Feroz:2013hea,Buchner:2014nha}.

Similarly, \Diver \cite{Martinez:2017lzg} requires the path to be given in \cq{DiverLib} in \cq{Setup}; other options are \cq{Convergence Threshold},\cq{Max Civilisations}, \cq{Number of parameters} and \cq{Verbose}. For the meaning of these see \cite{Martinez:2017lzg}.

%---------------------------------------------------------------------------------------------------------

\subsection{AL scans}
%\label{chap:bsmart:bsmart:scans:al}

In a likelihood-based approach, the goal is usually to reconstruct the likelihood distribution over the parameter space, typically to obtain a posterior distribution. In an MCMC approach, this means sampling such that the density of points is proportional to the likelihood in a given hypervolume;  more sophisticated treatments such as \MultiNest exist with the goal of inferring the posterior distribution more efficiently.  Alternatively, algorithms for finding the global maxima of the likelihood distribution exist, such as differential evolution or basin hopping. 

In many HEP applications, a likelihood function -- or at least, one that has a physical meaning -- is not available; or the likelihood distribution or most likely regions of parameter space may not be the questions of interest. For initial explorations of models, in fact, just the \emph{allowed} regions may be most interesting. Alternatively models may have large regions of mostly flat likelihood, for example when considering models of new physics without signals such as the muon $g-2$ or flavour anomalies. A classic problem of this kind might be finding the \emph{maximum} mass of dark matter, as we explored in \cite{Goodsell:2020rfu,Goodsell:2022beo}.

In \cite{Goodsell:2022beo} a class of scans was proposed using \emph{Active Learning} (AL) to tackle such problems where the aim is to find the decision boundary of a parameter space and the observables are a set of classifications (the point is either allowed or excluded). It can be especially useful when the ``oracle'' (in this case the BSM tools) are computationally expensive and an optimal choice of which points to test is desired. Starting with a small selection of random points, a neural network discriminator with multiple inputs but one output (having a sigmoid activation function) is trained to predict the decision boundary, and then the next set of points are chosen based on \emph{how uncertain} the network is, with the most interesting points being ones which maximise the function $y (1-y)$, where $y$ is the output of the discriminator. However, to train the network on batches of points, it is necessary to maximise the information gained by the whole batch, so we introduce a measure of diversity based on the distance between points. 

The full details of the algorithm are described in \cite{Goodsell:2022beo}; the results in that paper used our implementation in an early version of \BSMArt, and now anyone can use it for the first time. A typical input {\tt json} file (as used for \cite{Goodsell:2022beo}) containing the salient setttings for the scan would contain:
\begin{lstlisting}[language=MathIn,title=SMSQQ\_AL\_50k.json,basicstyle=\scriptsize]
  "Setup": {
        "RunName": "SMSQQ_AL",
        "Type": "AL",
        "StoreAllPoints":"False",
        "csv":"True",
        "Cores": 10,
        "Points":50000
    },
    "Networks": {
        "MLmodel": "SMSQQ_AL_50k",
        "HiddenSize": 100,
        "HiddenLayers": 3,
        "LearningRate": 0.001,
        "SGDmomentum": 0.1,
        "DSteps": 5000,
        "WeightDecay": 0.001,
        "Kinitial": 10000,
        "K": 500,
        "L": 100000,
        "FromGood": 0.2,
        "Epsilon": 0.8,
        "Diversity Alpha": 0.5,
        "FullTrain": 0,
        "AutoStop": "False"
    }
    "Variables": {
       "kappa":    { "RANGE": [1.5e5,1.85e5],
                       "VARIANCE": 2e4},
                     ...
      }
      "Observables": {
        "ms":      { "SLHA": ["MASS", [55]],
                      "TYPE":"OFF" },
        "Oh2":     { "SLHA": ["DARKMATTER", [1]],
                     "TYPE":"UPPER",
                     "MAX":0.112},

        "Vacstab": { "SLHA": ["VACUUMSTABILITY", [1]],
          "TYPE": "USER" },
        ...
    }
  \end{lstlisting}
All of the machine learning (ML) related settings are placed in \cq{Networks} whose settings we describe below.

In any given training round of the neural network, $L$ random points are generated, but without performing the time-consuming calculation of observables with HEPtools. 
These $L$ points are then passed to the neural network, which ranks them according to how sure it is about whether they are good or bad.
Only the $K$ points which the neural network is most uncertain about get passed on to the HEPtools.
$K$ is always smaller than $L$, and both values are set by the user.
The point selection algorithm is called \texttt{KfromL}.

Many parameter spaces are large, and only a small percentage of a random sample of points might be good. 
The AL scan in BSMArt therefore has the capability to propose more points from the vicinity of predetermined good points to the discriminator.
This ensures that the borders around small good areas are well explored.
This setting can be tuned by the user by changing the setting \texttt{FromGood}.

In order to not miss out on any good or bad areas, a proportion of random points bypasses the the \texttt{KfromL} algorithm.
This makes sure that the entire parameter space gets explored.

Training the discriminator can be tricky because it might get misled by a sample of points that diverges a lot from the average, or because it might diverge after some time, thus making more false estimates than earlier due to a too high learning rate or other suboptimally chosen settings.
To remedy this, the AL scan in BSMArt has several safeguard features. 
First, the learning rate can be decreased over time by choosing a value slightly below 1 for the setting \texttt{Epsilon}. 
This ensures that the network does not diverge because it has been misled by a new set of points.
Second, toggling the setting \texttt{AutoStop} on \texttt{True} ensures that a copy of the neural network is saved if the error rate drops below 5 percent, and that the scan stops running if the error rate subsequently rises above 20 percent.
The upside of this is that a fairly good neural network can be retained for further use, even when the capabilities of \texttt{Epsilon} do not suffice to save the training. 
The downside is that the target amount of points might not be reached.
Third, to avoid having to train on small and potentially erroneous sets of points, the setting \texttt{FullTrain} gives the user the liberty to choose how often they would like the discriminator to train on the full dataset rather than just the newest points. 
The downside is that training can take a considerable amount of time and computing power; the upside is that one might get a stabler scan.

The user must also choose some more settings for the neural network.
This includes the amount of nodes per hidden layer, \texttt{HiddenSize}, and the amount of hidden layers, \texttt{HiddenLayers}.
Generally speaking, the number of parameters of a neural network can be estimated to be around $S^L$, 
where $S$ corresponds to \texttt{HiddenSize} and $L$ to \texttt{HiddenLayers}. 
One should choose these settings such that the number of parameters is of the same order as the number of target points.

The \texttt{LearningRate} is a measure of how much information is retained each training cycle.
It is prudent, especially when starting with a new model, to keep this value low and keep the number of training cycles \texttt{DSteps} high to compensate for this.
This might consume more resources but reduces the probability of divergence during training.
Similarly, the setting \texttt{SGDmomentum} is a measure of the throughput of information due to the steepness of gradients. 
Generally speaking, one should choose smaller values for this when dealing with large networks, and one can afford to make it a bit larger with small networks.
Finally, \texttt{WeightDecay} ensures that weights decrease over time, therefore making the neural network smaller, more efficient, and speeding up training. 
It should be set higher the larger the network is.

All these settings are specified in the \texttt{Networks} block in a {JSON} file, where the scan \texttt{Type} is set to \texttt{AL}.
Examples JSON files are \texttt{InputExamples/SMSQQ/SMSQQ\_AL.json} and \\
\texttt{InputExamples/DiracGauginos/DiracGauginos\_AL.json}.
For the \cq{Variables}, only a \cq{RANGE} and \cq{VARIANCE} are required: the algorithm generates a proportion of new points based on jumps from the last points that were selected, with average jump distance given by the \cq{VARIANCE}. 
This is not as crucial as with an {MCMC} because the network can find interesting points by itself; %
however, in models where good points are scarce it can be useful to search for points in the vicinity of known good points. %

For \cq{Observables}, the possible settings for the classification are either \cq{OFF} (for those to be ignored); \cq{UPPER} for an upper limit where the observable is classed as excluded if it exceeds the \cq{MAX}; \cq{LOWER} for a lower limit where the  where the observable is classed as excluded if it is below the \cq{MIN}; \cq{USER} for a variable that can take values 0 or 1 (for example in the unitarity constraints); or if a \cq{RANGE}:[{\tt min},{\tt max}] is specifed the observable is classed as valid if it is between those ranges. The classification of a point as valid (1) is true if \emph{all} relevant observables are valid, and false (0) otherwise.

Finally, the AL scan provides a trained machine-learning model, which can be loaded and analysed, for example for plotting or retraining. The script {\tt BSMimport.py} is provided in the {\tt AuxFiles} directory which can perform this.

%\section{Plotting}
%\label{SEC:PLOTTING}

\subsection{Adding a new scan}
\label{sec:NewScans}

One of the principal goals of \BSMArt is to make implementing new scans as straighforward and transparent as possible. To create a scan with then name {\tt MyNewScan} the user just needs to \emph{either} place a file {\tt MyNewScan.py} in the {\tt bsmart/scans} directory \emph{or} in a subdirectory {\tt Scans} of the run directory from which the code is launched. This latter option is the preferred one for scans for a one-off project or those in development, because it does not interfere with the existing scripts and is more transparent for the user. It also allows additional user-written scripts to be loaded more easily. The new scan can then be used just by specifying the name under \cq{Type} in \cq{Setup}.

The structure of the file {\tt MyNewScan.py} should have the form:
\begin{lstlisting}[language=MathIn,title=MyNewScan.py,basicstyle=\scriptsize]
from core import Scan as Scan
import debug

class NewScan(Scan):
    """ The class should always be called NewScan """
 
    def __init__(self, inputs, log):
    """ Overload this for any initialisation required
        inputs is an ordered dictionary of the entire input json file (with some additional information added)    
    """
        Scan.__init__(self, inputs, log)
        
    """
    def initialise(self):
    # Overload this for initialisation that occurs *within* the __init__ routine (see core.py)
    # e.g. force tabbed output by
        self.runsettings.tabbed_output=True
    """ 
        
    def run(self):
        """ Required method that contains the actual running, does not return anything """
        ## code here
        
    def postprocess(self,Point, observables, data_point,temp_dir,log, lock=None):
        """ Routine for any post-processing on the point, e.g. likelihood calculation. Can return in any form desired """
        return ''
\end{lstlisting}
It is not necessary to overload/define the functions {\tt \_\_init\_\_, initialise, postprocess}: the user is only \emph{required} to define {\tt run}. To use the machinery of \BSMArt to run over all tools, the function {\tt self.RunManager.run\_batch(points)} can be used to run a list of points, where each point is a list of floats specifying the values of the variables. This will automatically use multiprocessing if specified in the {\tt json} file. The {\tt HEPRun.py} script and the classes contained within it actually controls all of the operations of running, and it can even be imported as a separate script on its own if the user wants to just use the running functions. For finer control, the scan has access to the individual {\tt CoreRuns} which run the operations on each core; e.g {\tt self.RunManager.CoreRuns[CoreNumber]}. These notably have the method {\tt self.RunManager.CoreRuns[CoreNumber].run\_point(Point, ID=None, lock=None)} where {\tt Point} is again just a list of floats, {\tt ID} is a number identifying the point (if desired) and {\tt lock} is for locking access to files during multicore running.

As with new tools, it is strongly encouraged to share new scans for dissemination with future versions of \BSMArt!

\section{Fitting}
\label{SEC:FITTING}

One optional feature of \BSMArt is the possibility of marginalising over certain input variables. This is intended for cases when the value of an observable is well known but the inverse function relating it to input variables is complicated: for example trading the Higgs mass for one or more parameters (stop masses or trilinear couplings in supersymmetric theories) and/or matching the dark matter relic density. In principle any number of variables can be traded for any number of observables; this is exemplified in {\tt Fitting\_MSSM\_mh.json}:
\begin{lstlisting}[language=MathIn,title=Fitting\_MSSM\_mh.json,basicstyle=\scriptsize]
   "Codes":{
        "SPheno":{
             "Command": "~/BSMArt/SPheno-4.0.5/bin/SPhenoMSSM",
             "InputFile": "fast.in",
            "Run": "True",
	    "Observables": {"mh" : { "SLHA": ["MASS", [25]], "MEAN": 125.0, "VARIANCE": 3.0, "SCALING":"GAUSS"}}
        }
    },
   "Variables":{ "m12" : { "RANGE": [200,4000]},
		  "A0" : { "RANGE": [-6000,6000]}
		},
    "Fitting": {
	"Variables": { "m0" : {"RANGE": [200,6000]} },
	"Observables": {"mh" :{"MEAN": 125,"VARIANCE":0.5}},
	"Options" : {"eps": 0.1, "ftol" : 1.0, "disp" : "False"}
    }
\end{lstlisting}
The idea is that the variables that appear in \cq{Fitting} \emph{do not appear} in the usual \cq{Variables} list: they do not appear in the list of variables to be scanned over, and this can therefore be used with \emph{any} scan type. When the \cq{Fitting} dictionary is present, for the case of one variable being traded for one observable, a modified root-finding algorithm is used to attempt to match the two with a mean and error given by the values given in \cq{Fitting}:\cq{Observables} (and options to be passed to the optimisation algorithm are given there).

On the other hand, if there is more than one variable or observable to be fit, the code constructs the function
\begin{equation}
  \chi^2 = \sum_{i} \frac{(y_i - m_i)^2}{2 \sigma_i^2}
\end{equation}
for observables $y_i$  with mean $m_i$ and variance $\sigma_i$. This is then minimised over the specified variables within the specified ranges using the {\tt python} {\tt scipy.optimize.minimise} function. An additional option \cq{Method} can be specified within \cq{Fitting} to determine the algorithm, e.g. \cq{Nelder-Mead}, \cq{BFGS} etc; see the {\tt scipy} documentation. Since these algorithms mostly assume an infinite range, the variables $x$ seen by the optimiser are scaled to the range $[a,b]$ using
\begin{align}
v =&  \frac{a+b}{2} + \frac{(b-a)}{\pi} \tan^{-1} (x),
\end{align}
just as in {\tt SSP}. But it should be stressed this is only applied to the multivariate case; single variable scans require no rescaling because Brent's method is used after an initial search.

The code has an additional feature that, during the fitting stage, \BSMArt only runs the codes up to the last one necessary for the optimisation. For example, if the first step is to run \SPheno and the Higgs mass is searched for in exchange for the Higgs quartic coupling, then this is performed using only \SPheno; the full code chain is only run after the optimal quartic coupling has been determined. This can save significant computation time. On the other hand, the fitting does still require several evaluations of \SPheno, and other techniques (e.g. machine learning to learn the Higgs mass) may be more efficient. 

\section{Conclusions and future directions}
\label{SEC:CONCLUSIONS}

With the release of \BSMArt, much of the effort of exploring the parameter spaces of new models can be automated. It includes highly flexible and general algorithms with many convenience functions. Scans can be set up in a matter of minutes by modifying the {\tt json} file generated by {\tt prepareModel.py} and the {\tt SLHA} file generated by \SARAH. It also releases for the first time the code for the Active Learning scan \cite{Goodsell:2022beo}.

Due to its flexibility there are many possible avenues for future improvements. In the next version it is intended to include an interface to use {\tt CalcHEP}\cite{Belyaev:2012qa} for cross-section generation and {\tt SmodelS} \cite{Alguero:2021dig} for fast collider limits. A tool to interface with {\tt smelli} \cite{Stangl:2020lbh} is in preparation; {\tt Lilith} \cite{Bernon:2015hsa,Bertrand:2020lyb} and \madanalysis \cite{Conte:2012fm,Conte:2014zja,Conte:2018vmg,Araz:2021akd} are envisaged. In addition, it would be highly useful to interface to other statistics packages, for computing collider limits, or for computing likelihoods based on collected data; even if this is not the primary goal for \BSMArt, it is sufficiently powerful to handle such tasks.  

On the other hand, the greatest scope for new developments is on applications of machine learning -- or alternative algorithms -- to parameter space exploration, in the form of new types of scans, which is where \BSMArt is strongest. The possibilities in this direction are endless.

%%%%%%%%%%%%%%%%%%%%%%%%%%%%%%%%%%%%%%%%%%%%%%%%%%%%%%%%%%%%%%%%%%%%%%%%%%%%%%%%%%%%%%%%%%%%%%%%%%%%%%%%%

\section*{Acknowledgements}

MDG acknowledges support from the grants
\mbox{``HiggsAutomator''} and \mbox{``DMwithLLPatLHC''} of the Agence Nationale de la Recherche
(ANR) (ANR-15-CE31-0002), (ANR-21-CE31-0013). We thank Humberto Reyes-Gonz\'alez for sharing some code at an early stage. We thank Florian Staub and Luc Darm\'e for discussions at an early stage of this project; Farid Ibrahimov and Kartik Sharma for providing feedback on preliminary versions; and Werner Porod for comments on the draft.

\bibliographystyle{h-physrev}
\bibliography{lit}

\begin{thebibliography}{10}

\bibitem{Allanach:2008qq}
B.~Allanach, \emph{et~al.},
\newblock ``{SUSY Les Houches Accord 2}'',
\newblock
  \href{http://dx.doi.org/10.1016/j.cpc.2008.08.004}{Comput.Phys.Commun.\,\textbf{180},\,8\,(2009)},
  \href{http://arxiv.org/abs/0801.0045}{arXiv:0801.0045}.

\bibitem{Buchmueller:2007zk}
O.~Buchmueller, R.~Cavanaugh, A.~De~Roeck, S.~Heinemeyer, G.~Isidori,
  P.~Paradisi, F.~J. Ronga, A.~M. Weber, G.~Weiglein,
\newblock ``{Prediction for the Lightest Higgs Boson Mass in the CMSSM using
  Indirect Experimental Constraints}'',
\newblock \href{http://dx.doi.org/10.1016/j.physletb.2007.09.058}{Phys. Lett.
  B\,\textbf{657},\,87\,(2007)},
  \href{http://arxiv.org/abs/0707.3447}{arXiv:0707.3447}.

\bibitem{Bagnaschi:2019djj}
E.~Bagnaschi, \emph{et~al.},
\newblock ``{Global Analysis of Dark Matter Simplified Models with Leptophobic
  Spin-One Mediators using MasterCode}'',
\newblock \href{http://dx.doi.org/10.1140/epjc/s10052-019-7382-3}{Eur. Phys. J.
  C\,\textbf{79},\,895\,(2019)},
  \href{http://arxiv.org/abs/1905.00892}{arXiv:1905.00892}.

\bibitem{Darme:2019wpd}
L.~Darm\'e, A.~Hryczuk, D.~Karamitros, L.~Roszkowski,
\newblock ``{Forbidden frozen-in dark matter}'',
\newblock
  \href{http://dx.doi.org/10.1007/JHEP11(2019)159}{JHEP\,\textbf{11},\,159\,(2019)},
  \href{http://arxiv.org/abs/1908.05685}{arXiv:1908.05685}.

\bibitem{Ahmed:2022jlo}
W.~Ahmed, M.~Goodsell, S.~Munir,
\newblock ``{Dark matter in the CP-violating NMSSM}'',
\newblock \href{http://dx.doi.org/10.1140/epjc/s10052-022-10449-z}{Eur. Phys.
  J. C\,\textbf{82},\,539\,(2022)},
  \href{http://arxiv.org/abs/2201.10628}{arXiv:2201.10628}.

\bibitem{Staub:2011dp}
F.~Staub, T.~Ohl, W.~Porod, C.~Speckner,
\newblock ``{A Tool Box for Implementing Supersymmetric Models}'',
\newblock \href{http://dx.doi.org/10.1016/j.cpc.2012.04.013}{Comput. Phys.
  Commun.\,\textbf{183},\,2165\,(2012)},
  \href{http://arxiv.org/abs/1109.5147}{arXiv:1109.5147}.

\bibitem{GAMBIT:2017yxo}
GAMBIT, P.~Athron, \emph{et~al.},
\newblock ``{GAMBIT: The Global and Modular Beyond-the-Standard-Model Inference
  Tool}'',
\newblock \href{http://dx.doi.org/10.1140/epjc/s10052-017-5321-8}{Eur. Phys. J.
  C\,\textbf{77},\,784\,(2017)},
  \href{http://arxiv.org/abs/1705.07908}{arXiv:1705.07908},
\newblock [Addendum: Eur.Phys.J.C 78, 98 (2018)].

\bibitem{Kvellestad:2019vxm}
A.~Kvellestad, P.~Scott, M.~White,
\newblock ``{GAMBIT and its Application in the Search for Physics Beyond the
  Standard Model}'',
\newblock \href{http://dx.doi.org/10.1016/j.ppnp.2020.103769}{\,\,(2019)},
  \href{http://arxiv.org/abs/1912.04079}{arXiv:1912.04079}.

\bibitem{Bloor:2021gtp}
S.~Bloor, T.~E. Gonzalo, P.~Scott, C.~Chang, A.~Raklev, J.~E. Camargo-Molina,
  A.~Kvellestad, J.~J. Renk, P.~Athron, C.~Bal\'azs,
\newblock ``{The GAMBIT Universal Model Machine: from Lagrangians to
  likelihoods}'',
\newblock \href{http://dx.doi.org/10.1140/epjc/s10052-021-09828-9}{Eur. Phys.
  J. C\,\textbf{81},\,1103\,(2021)},
  \href{http://arxiv.org/abs/2107.00030}{arXiv:2107.00030}.

\bibitem{Goodsell:2022beo}
M.~D. Goodsell, A.~Joury,
\newblock ``{Active learning BSM parameter spaces}'',
\newblock \,\,(2022), \href{http://arxiv.org/abs/2204.13950}{arXiv:2204.13950}.

\bibitem{Staub:2019xhl}
F.~Staub,
\newblock ``{xBIT: an easy to use scanning tool with machine learning
  abilities}'',
\newblock \,\,(2019), \href{http://arxiv.org/abs/1906.03277}{arXiv:1906.03277}.

\bibitem{Staub:2013tta}
F.~Staub,
\newblock ``{SARAH 4 : A tool for (not only SUSY) model builders}'',
\newblock \href{http://dx.doi.org/10.1016/j.cpc.2014.02.018}{Comput. Phys.
  Commun.\,\textbf{185},\,1773\,(2014)},
  \href{http://arxiv.org/abs/1309.7223}{arXiv:1309.7223}.

\bibitem{Goodsell:2017pdq}
M.~D. Goodsell, S.~Liebler, F.~Staub,
\newblock ``{Generic calculation of two-body partial decay widths at the full
  one-loop level}'',
\newblock \href{http://dx.doi.org/10.1140/epjc/s10052-017-5259-x}{Eur. Phys. J.
  C\,\textbf{77},\,758\,(2017)},
  \href{http://arxiv.org/abs/1703.09237}{arXiv:1703.09237}.

\bibitem{Braathen:2017izn}
J.~Braathen, M.~D. Goodsell, F.~Staub,
\newblock ``{Supersymmetric and non-supersymmetric models without catastrophic
  Goldstone bosons}'',
\newblock \href{http://dx.doi.org/10.1140/epjc/s10052-017-5303-x}{Eur. Phys. J.
  C\,\textbf{77},\,757\,(2017)},
  \href{http://arxiv.org/abs/1706.05372}{arXiv:1706.05372}.

\bibitem{Goodsell:2018tti}
M.~D. Goodsell, F.~Staub,
\newblock ``{Unitarity constraints on general scalar couplings with SARAH}'',
\newblock \href{http://dx.doi.org/10.1140/epjc/s10052-018-6127-z}{Eur. Phys.
  J.\,\textbf{C78},\,649\,(2018)},
  \href{http://arxiv.org/abs/1805.07306}{arXiv:1805.07306}.

\bibitem{Goodsell:2020rfu}
M.~D. Goodsell, R.~Moutafis,
\newblock ``{How heavy can dark matter be? Constraining colourful unitarity
  with SARAH}'',
\newblock \href{http://dx.doi.org/10.1140/epjc/s10052-021-09597-5}{Eur. Phys.
  J. C\,\textbf{81},\,808\,(2021)},
  \href{http://arxiv.org/abs/2012.09022}{arXiv:2012.09022}.

\bibitem{Benakli:2022gjn}
K.~Benakli, M.~Goodsell, W.~Ke, P.~Slavich,
\newblock ``{W boson mass in minimal Dirac gaugino scenarios}'',
\newblock \,\,(2022), \href{http://arxiv.org/abs/2208.05867}{arXiv:2208.05867}.

\bibitem{Bagnaschi:2022zvd}
E.~Bagnaschi, M.~Goodsell, P.~Slavich,
\newblock ``{Higgs-mass prediction in the NMSSM with heavy BSM particles}'',
\newblock \href{http://dx.doi.org/10.1140/epjc/s10052-022-10810-2}{Eur. Phys.
  J. C\,\textbf{82},\,853\,(2022)},
  \href{http://arxiv.org/abs/2206.04618}{arXiv:2206.04618}.

\bibitem{Porod:2011nf}
W.~Porod, F.~Staub,
\newblock ``{SPheno 3.1: Extensions including flavour, CP-phases and models
  beyond the MSSM}'',
\newblock \,\,(2011), \href{http://arxiv.org/abs/1104.1573}{arXiv:1104.1573}.

\bibitem{Goodsell:2021iwc}
M.~D. Goodsell, L.~Priya,
\newblock ``{Long dead winos}'',
\newblock \href{http://dx.doi.org/10.1140/epjc/s10052-022-10188-1}{Eur. Phys.
  J. C\,\textbf{82},\,235\,(2022)},
  \href{http://arxiv.org/abs/2106.08815}{arXiv:2106.08815}.

\bibitem{Porod:2014xia}
W.~Porod, F.~Staub, A.~Vicente,
\newblock ``{A Flavor Kit for BSM models}'',
\newblock \href{http://dx.doi.org/10.1140/epjc/s10052-014-2992-2}{Eur. Phys. J.
  C\,\textbf{74},\,2992\,(2014)},
  \href{http://arxiv.org/abs/1405.1434}{arXiv:1405.1434}.

\bibitem{Staub:2018rih}
F.~Staub,
\newblock ``{xSLHA: an Les Houches Accord reader for Python and Mathematica}'',
\newblock \href{http://dx.doi.org/10.1016/j.cpc.2019.03.013}{Comput. Phys.
  Commun.\,\textbf{241},\,132\,(2019)},
  \href{http://arxiv.org/abs/1812.04655}{arXiv:1812.04655}.

\bibitem{Buckley:2013jua}
A.~Buckley,
\newblock ``{PySLHA: a Pythonic interface to SUSY Les Houches Accord data}'',
\newblock \href{http://dx.doi.org/10.1140/epjc/s10052-015-3638-8}{Eur. Phys. J.
  C\,\textbf{75},\,467\,(2015)},
  \href{http://arxiv.org/abs/1305.4194}{arXiv:1305.4194}.

\bibitem{Porod:2003um}
W.~Porod,
\newblock ``{SPheno, a program for calculating supersymmetric spectra, SUSY
  particle decays and SUSY particle production at e+ e- colliders}'',
\newblock
  \href{http://dx.doi.org/10.1016/S0010-4655(03)00222-4}{Comput.Phys.Commun.\,\textbf{153},\,275\,(2003)},
  \href{http://arxiv.org/abs/hep-ph/0301101}{arXiv:hep-ph/0301101}.

\bibitem{Staub:2008uz}
F.~Staub,
\newblock ``{SARAH}'',
\newblock \,\,(2008), \href{http://arxiv.org/abs/0806.0538}{arXiv:0806.0538}.

\bibitem{Goodsell:2015ira}
M.~Goodsell, K.~Nickel, F.~Staub,
\newblock ``{Generic two-loop Higgs mass calculation from a diagrammatic
  approach}'',
\newblock \href{http://dx.doi.org/10.1140/epjc/s10052-015-3494-6}{Eur. Phys. J.
  C\,\textbf{75},\,290\,(2015)},
  \href{http://arxiv.org/abs/1503.03098}{arXiv:1503.03098}.

\bibitem{Belanger:2007zz}
G.~Belanger, F.~Boudjema, A.~Pukhov, A.~Semenov,
\newblock ``{micrOMEGAs 2.0.7: A program to calculate the relic density of dark
  matter in a generic model}'',
\newblock
  \href{http://dx.doi.org/10.1016/j.cpc.2007.08.002}{Comput.Phys.Commun.\,\textbf{177},\,894\,(2007)}.

\bibitem{Belanger:2010pz}
G.~Belanger, F.~Boudjema, A.~Pukhov, A.~Semenov,
\newblock ``{micrOMEGAs: A Tool for dark matter studies}'',
\newblock \,\,(2010), \href{http://arxiv.org/abs/1005.4133}{arXiv:1005.4133}.

\bibitem{Belanger:2018ccd}
G.~B\'elanger, F.~Boudjema, A.~Goudelis, A.~Pukhov, B.~Zaldivar,
\newblock ``{micrOMEGAs5.0 : Freeze-in}'',
\newblock \href{http://dx.doi.org/10.1016/j.cpc.2018.04.027}{Comput. Phys.
  Commun.\,\textbf{231},\,173\,(2018)},
  \href{http://arxiv.org/abs/1801.03509}{arXiv:1801.03509}.

\bibitem{Bechtle:2008jh}
P.~Bechtle, O.~Brein, S.~Heinemeyer, G.~Weiglein, K.~E. Williams,
\newblock ``{HiggsBounds: Confronting Arbitrary Higgs Sectors with Exclusion
  Bounds from LEP and the Tevatron}'',
\newblock \href{http://dx.doi.org/10.1016/j.cpc.2009.09.003}{Comput. Phys.
  Commun.\,\textbf{181},\,138\,(2010)},
  \href{http://arxiv.org/abs/0811.4169}{arXiv:0811.4169}.

\bibitem{Bechtle:2011sb}
P.~Bechtle, O.~Brein, S.~Heinemeyer, G.~Weiglein, K.~E. Williams,
\newblock ``{HiggsBounds 2.0.0: Confronting Neutral and Charged Higgs Sector
  Predictions with Exclusion Bounds from LEP and the Tevatron}'',
\newblock \href{http://dx.doi.org/10.1016/j.cpc.2011.07.015}{Comput. Phys.
  Commun.\,\textbf{182},\,2605\,(2011)},
  \href{http://arxiv.org/abs/1102.1898}{arXiv:1102.1898}.

\bibitem{Bechtle:2013wla}
P.~Bechtle, O.~Brein, S.~Heinemeyer, O.~St\r{a}l, T.~Stefaniak, G.~Weiglein,
  K.~E. Williams,
\newblock ``{$\mathsf{HiggsBounds}-4$: Improved Tests of Extended Higgs Sectors
  against Exclusion Bounds from LEP, the Tevatron and the LHC}'',
\newblock \href{http://dx.doi.org/10.1140/epjc/s10052-013-2693-2}{Eur. Phys. J.
  C\,\textbf{74},\,2693\,(2014)},
  \href{http://arxiv.org/abs/1311.0055}{arXiv:1311.0055}.

\bibitem{Bechtle:2020pkv}
P.~Bechtle, D.~Dercks, S.~Heinemeyer, T.~Klingl, T.~Stefaniak, G.~Weiglein,
  J.~Wittbrodt,
\newblock ``{HiggsBounds-5: Testing Higgs Sectors in the LHC 13 TeV Era}'',
\newblock \href{http://dx.doi.org/10.1140/epjc/s10052-020-08557-9}{Eur. Phys.
  J. C\,\textbf{80},\,1211\,(2020)},
  \href{http://arxiv.org/abs/2006.06007}{arXiv:2006.06007}.

\bibitem{Bechtle:2013xfa}
P.~Bechtle, S.~Heinemeyer, O.~St\r{a}l, T.~Stefaniak, G.~Weiglein,
\newblock ``{$HiggsSignals$: Confronting arbitrary Higgs sectors with
  measurements at the Tevatron and the LHC}'',
\newblock \href{http://dx.doi.org/10.1140/epjc/s10052-013-2711-4}{Eur. Phys. J.
  C\,\textbf{74},\,2711\,(2014)},
  \href{http://arxiv.org/abs/1305.1933}{arXiv:1305.1933}.

\bibitem{Bechtle:2020uwn}
P.~Bechtle, S.~Heinemeyer, T.~Klingl, T.~Stefaniak, G.~Weiglein, J.~Wittbrodt,
\newblock ``{HiggsSignals-2: Probing new physics with precision Higgs
  measurements in the LHC 13 TeV era}'',
\newblock \href{http://dx.doi.org/10.1140/epjc/s10052-021-08942-y}{Eur. Phys.
  J. C\,\textbf{81},\,145\,(2021)},
  \href{http://arxiv.org/abs/2012.09197}{arXiv:2012.09197}.

\bibitem{Bahl:2022igd}
H.~Bahl, T.~Biek\"otter, S.~Heinemeyer, C.~Li, S.~Paasch, G.~Weiglein,
  J.~Wittbrodt,
\newblock ``{HiggsTools: BSM scalar phenomenology with new versions of
  HiggsBounds and HiggsSignals}'',
\newblock \,\,(2022), \href{http://arxiv.org/abs/2210.09332}{arXiv:2210.09332}.

\bibitem{Camargo-Molina:2013qva}
J.~E. Camargo-Molina, B.~O'Leary, W.~Porod, F.~Staub,
\newblock ``{$\mathbf{Vevacious}$: A Tool For Finding The Global Minima Of
  One-Loop Effective Potentials With Many Scalars}'',
\newblock \href{http://dx.doi.org/10.1140/epjc/s10052-013-2588-2}{Eur. Phys. J.
  C\,\textbf{73},\,2588\,(2013)},
  \href{http://arxiv.org/abs/1307.1477}{arXiv:1307.1477}.

\bibitem{Alwall:2011uj}
J.~Alwall, M.~Herquet, F.~Maltoni, O.~Mattelaer, T.~Stelzer,
\newblock ``{MadGraph 5 : Going Beyond}'',
\newblock
  \href{http://dx.doi.org/10.1007/JHEP06(2011)128}{JHEP\,\textbf{06},\,128\,(2011)},
  \href{http://arxiv.org/abs/1106.0522}{arXiv:1106.0522}.

\bibitem{Straub:2018kue}
D.~M. Straub,
\newblock ``{flavio: a Python package for flavour and precision phenomenology
  in the Standard Model and beyond}'',
\newblock \,\,(2018), \href{http://arxiv.org/abs/1810.08132}{arXiv:1810.08132}.

\bibitem{Aebischer:2017ugx}
J.~Aebischer, \emph{et~al.},
\newblock ``{WCxf: an exchange format for Wilson coefficients beyond the
  Standard Model}'',
\newblock \href{http://dx.doi.org/10.1016/j.cpc.2018.05.022}{Comput. Phys.
  Commun.\,\textbf{232},\,71\,(2018)},
  \href{http://arxiv.org/abs/1712.05298}{arXiv:1712.05298}.

\bibitem{Feroz:2008xx}
F.~Feroz, M.~P. Hobson, M.~Bridges,
\newblock ``{MultiNest: an efficient and robust Bayesian inference tool for
  cosmology and particle physics}'',
\newblock \href{http://dx.doi.org/10.1111/j.1365-2966.2009.14548.x}{Mon. Not.
  Roy. Astron. Soc.\,\textbf{398},\,1601\,(2009)},
  \href{http://arxiv.org/abs/0809.3437}{arXiv:0809.3437}.

\bibitem{Feroz:2013hea}
F.~Feroz, M.~P. Hobson, E.~Cameron, A.~N. Pettitt,
\newblock ``{Importance Nested Sampling and the MultiNest Algorithm}'',
\newblock \href{http://dx.doi.org/10.21105/astro.1306.2144}{Open J.
  Astrophys.\,\textbf{2},\,10\,(2019)},
  \href{http://arxiv.org/abs/1306.2144}{arXiv:1306.2144}.

\bibitem{Buchner:2014nha}
J.~Buchner, A.~Georgakakis, K.~Nandra, L.~Hsu, C.~Rangel, M.~Brightman,
  A.~Merloni, M.~Salvato, J.~Donley, D.~Kocevski,
\newblock ``{X-ray spectral modelling of the AGN obscuring region in the CDFS:
  Bayesian model selection and catalogue}'',
\newblock \href{http://dx.doi.org/10.1051/0004-6361/201322971}{Astron.
  Astrophys.\,\textbf{564},\,A125\,(2014)},
  \href{http://arxiv.org/abs/1402.0004}{arXiv:1402.0004}.

\bibitem{Martinez:2017lzg}
GAMBIT, G.~D. Martinez, J.~McKay, B.~Farmer, P.~Scott, E.~Roebber, A.~Putze,
  J.~Conrad,
\newblock ``{Comparison of statistical sampling methods with ScannerBit, the
  GAMBIT scanning module}'',
\newblock \href{http://dx.doi.org/10.1140/epjc/s10052-017-5274-y}{Eur. Phys. J.
  C\,\textbf{77},\,761\,(2017)},
  \href{http://arxiv.org/abs/1705.07959}{arXiv:1705.07959}.

\bibitem{Brivio:2019ius}
I.~Brivio, S.~Bruggisser, F.~Maltoni, R.~Moutafis, T.~Plehn, E.~Vryonidou,
  S.~Westhoff, C.~Zhang,
\newblock ``{O new physics, where art thou? A global search in the top
  sector}'',
\newblock
  \href{http://dx.doi.org/10.1007/JHEP02(2020)131}{JHEP\,\textbf{02},\,131\,(2020)},
  \href{http://arxiv.org/abs/1910.03606}{arXiv:1910.03606}.

\bibitem{Belyaev:2012qa}
A.~Belyaev, N.~D. Christensen, A.~Pukhov,
\newblock ``{CalcHEP 3.4 for collider physics within and beyond the Standard
  Model}'',
\newblock \href{http://dx.doi.org/10.1016/j.cpc.2013.01.014}{Comput. Phys.
  Commun.\,\textbf{184},\,1729\,(2013)},
  \href{http://arxiv.org/abs/1207.6082}{arXiv:1207.6082}.

\bibitem{Alguero:2021dig}
G.~Alguero, J.~Heisig, C.~Khosa, S.~Kraml, S.~Kulkarni, A.~Lessa,
  H.~Reyes-Gonz\'alez, W.~Waltenberger, A.~Wongel,
\newblock ``{Constraining new physics with SModelS version 2}'',
\newblock \,\,(2021), \href{http://arxiv.org/abs/2112.00769}{arXiv:2112.00769}.

\bibitem{Stangl:2020lbh}
P.~Stangl,
\newblock ``{smelli \textendash{} the SMEFT Likelihood}'',
\newblock
  \href{http://dx.doi.org/10.22323/1.392.0035}{PoS\,\textbf{TOOLS2020},\,035\,(2021)},
  \href{http://arxiv.org/abs/2012.12211}{arXiv:2012.12211}.

\bibitem{Bernon:2015hsa}
J.~Bernon, B.~Dumont,
\newblock ``{Lilith: a tool for constraining new physics from Higgs
  measurements}'',
\newblock \href{http://dx.doi.org/10.1140/epjc/s10052-015-3645-9}{Eur. Phys. J.
  C\,\textbf{75},\,440\,(2015)},
  \href{http://arxiv.org/abs/1502.04138}{arXiv:1502.04138}.

\bibitem{Bertrand:2020lyb}
M.~Bertrand, S.~Kraml, T.~Q. Loc, D.~T. Nhung, L.~D. Ninh,
\newblock ``{Constraining new physics from Higgs measurements with Lilith-2}'',
\newblock
  \href{http://dx.doi.org/10.22323/1.392.0040}{PoS\,\textbf{TOOLS2020},\,040\,(2021)},
  \href{http://arxiv.org/abs/2012.11408}{arXiv:2012.11408}.

\bibitem{Conte:2012fm}
E.~Conte, B.~Fuks, G.~Serret,
\newblock ``{MadAnalysis 5, A User-Friendly Framework for Collider
  Phenomenology}'',
\newblock \href{http://dx.doi.org/10.1016/j.cpc.2012.09.009}{Comput. Phys.
  Commun.\,\textbf{184},\,222\,(2013)},
  \href{http://arxiv.org/abs/1206.1599}{arXiv:1206.1599}.

\bibitem{Conte:2014zja}
E.~Conte, B.~Dumont, B.~Fuks, C.~Wymant,
\newblock ``{Designing and recasting LHC analyses with MadAnalysis 5}'',
\newblock \href{http://dx.doi.org/10.1140/epjc/s10052-014-3103-0}{Eur. Phys. J.
  C\,\textbf{74},\,3103\,(2014)},
  \href{http://arxiv.org/abs/1405.3982}{arXiv:1405.3982}.

\bibitem{Conte:2018vmg}
E.~Conte, B.~Fuks,
\newblock ``{Confronting new physics theories to LHC data with MADANALYSIS
  5}'',
\newblock \href{http://dx.doi.org/10.1142/S0217751X18300272}{Int. J. Mod. Phys.
  A\,\textbf{33},\,1830027\,(2018)},
  \href{http://arxiv.org/abs/1808.00480}{arXiv:1808.00480}.

\bibitem{Araz:2021akd}
J.~Y. Araz, B.~Fuks, M.~D. Goodsell, M.~Utsch,
\newblock ``{Recasting LHC searches for long-lived particles with
  MadAnalysis~5}'',
\newblock \href{http://dx.doi.org/10.1140/epjc/s10052-022-10511-w}{Eur. Phys.
  J. C\,\textbf{82},\,597\,(2022)},
  \href{http://arxiv.org/abs/2112.05163}{arXiv:2112.05163}.

\end{thebibliography}

\end{document}